\pdfoutput=1
\documentclass[11pt]{article}

\usepackage[letterpaper,margin=1in]{geometry}
\usepackage[T1]{fontenc}
\usepackage[utf8]{inputenc}
\usepackage{authblk}
\usepackage[numbers,sort&compress]{natbib}
\usepackage{microtype}
\usepackage{hyperref}

\usepackage{booktabs}
\usepackage{array}
\usepackage{multirow}
\usepackage{amsmath}
\usepackage{amssymb}
\usepackage{graphicx}
\usepackage{tikz}
\usepackage{xspace}
\usepackage{paralist}
\usepackage{algorithm}
\usepackage{algpseudocode}
\usepackage[capitalise,noabbrev]{cleveref}
\usepackage{xcolor}

\newcommand{\sys}{\textsc{DreamTest}\xspace}
\newcommand{\indago}{\textsc{Indago}\xspace}
\newcommand{\nexus}{\textsc{Indago-Nexus}\xspace}
\newcommand{\mdpfuzz}{\textsc{MDPFuzz}\xspace}
\newcommand{\samota}{\textsc{SAMOTA}\xspace}
\newcommand{\pfail}{s(c)}

\newcommand{\auprc}{AUPRC}
\newcommand{\auroc}{AUROC}

\usepackage{framed}
\newcommand{\rqconclusion}[2]{%
  \begin{framed}\noindent\textbf{Answer to #1.} #2\end{framed}}

\begin{document}

\title{\sys: World-Model Surrogates for Search-Based Testing of Deep Reinforcement Learning Agents}

\author[1]{Qinghua Xu}
\author[1]{Guancheng Wang}
\author[1]{Boxi Yu}
\author[1]{Liting Lin}
\author[1,2]{Lionel Briand}
\affil[1]{Lero Research Centre, University of Limerick, Ireland}
\affil[2]{University of Ottawa, Canada}
\date{}

\maketitle

\begin{abstract}
Testing deep reinforcement learning (DRL) agents in cyber-physical systems aims to uncover diverse failures before deployment, but each execution can be expensive. Surrogate-assisted testing reduces this cost by learning to predict which test configurations are likely to fail. Prior surrogates treat the system as a black box and predict pass/fail outcomes directly; we instead model how a test unfolds and estimate failure from an imagined episode.

We introduce \sys, a world-model surrogate for testing DRL agents. \sys adapts a recurrent state-space model to learn agent behaviour and environment dynamics from the agent's training log. Given a candidate configuration, imagined rollouts produce a failure score that guides search without executing every candidate in a simulator or real system.

We evaluate \sys for failure prediction, test generation, and failure diversity on Parking, Humanoid, and DonkeyCar. Mean area under the precision--recall curve (\auprc) exceeds the strongest baseline by 97\%, 12\%, and 39\%,  respectively, and gains on five out-of-distribution test sets reach 145\%, 29\%, and 44\%. Under the same simulator-validation budget, the best ``\textit{\sys + search}'' combinations find 29\%, 22\%, and 79\% more novel failures on average. Across clusterings with $k= 2,\dots ,40$, failures generated with \sys cover the most behavioural clusters for almost all $k$, indicating that \sys consistently discovers behaviourally diverse failures.
\end{abstract}


\section{Introduction}
\label{sec:intro}

Deep reinforcement learning (DRL) agents increasingly control cyber-physical systems (CPS), from vehicle parking and vision-based driving to humanoid control~\cite{haarnoja2018sac,kuznetsov2020tqc,biagiola2024indago}. Testing these agents aims to expose as many failures as possible so that weaknesses can be addressed before deployment. Each test specifies an \textit{environment configuration}, such as a parking layout, an initial joint state, or a road geometry. Exhaustive testing is infeasible because configuration spaces are too large to enumerate, and validating a single generated test can require a simulator run lasting tens of seconds~\cite{uesato2019rigorous,biagiola2024indago}. The challenge is therefore to discover many failures, spanning a broad range of behaviours, under a limited simulator-validation budget. 

A common solution is \textit{search-based testing}, which frames failure discovery as an optimisation problem: feedback from executed candidates guides the selection of subsequent configurations. For learning-enabled CPS, evolutionary search has been used to find scenarios that expose unsafe behaviour~\cite{abdessalem2016ase,abdessalem2018icse}. However, executing every evaluated candidate in the simulator remains costly. Learned surrogate models provide an alternative and are increasingly used to support test oracles, evolving digital twins, and anomaly detection across CPS and software systems~\cite{wang2025qurebot,xu2024ppt,xu2023kddt,hadadi2026flexlog}. In search-based testing, these models replace expensive evaluations with inexpensive predictions, reserving simulator validation for the most promising tests~\cite{jin2011surrogate,haq2022samota}. For instance, the Surrogate-Assisted Many-Objective Testing Approach (\samota) uses classical surrogate families, including Kriging, polynomial, and radial basis function models, to guide the testing of driving systems~\cite{haq2022samota}.

For DRL agents, training itself provides a rich data source for learning such surrogates. Uesato et al.\ learn a failure predictor from the agent's training history and use it to prioritise configurations for testing~\cite{uesato2019rigorous}. \indago~\cite{biagiola2024indago} develops this idea into a surrogate-assisted pipeline: it logs each episode's configuration and pass/fail outcome, fits a classifier to estimate failure probability, and searches for configurations that maximise this estimate. \nexus~\cite{bartlett2025nexus} extends this line with many-objective search to promote failure diversity.

However, these surrogates are discriminative: they map configurations to pass/fail labels while treating CPS operation as a black box. In contrast, a \textit{world model} is a learned predictive model of system dynamics: given a state and an action, it predicts the next state and related signals such as reward and termination. Rolling this model forward simulates an episode trajectory without executing the system in the environment. This idea underlies World Models and Dreamer~\cite{ha2018worldmodels,hafner2020dreamer}. For testing, a world model trained on an agent's training log could simulate the episode induced by a candidate configuration and estimate whether it ends in failure.

To that end, we introduce \sys, a surrogate that learns a world model from a DRL agent's training log and scores configurations by simulating how their episodes unfold inside the learned model. We adapt a recurrent state-space model (RSSM)~\cite{hafner2019planet} for testing and train it on per-step interactions from the training log. The model predicts observations, rewards, the agent's next action, and termination and failure. Given a candidate configuration, \sys simulates $K$ possible episode trajectories in parallel (\textit{imagined rollouts}). After the rollouts terminate, \sys computes a failure score $\pfail$ by combining the predicted failures, weighted by predicted termination probabilities along each trajectory.

We evaluate \sys on three DRL benchmarks: \textit{Parking} (vehicle parking), \textit{Humanoid} (humanoid control), and \textit{DonkeyCar} (self-driving). For direct failure prediction, \sys achieves the highest mean area under the precision--recall curve (\auprc) throughout, exceeding the strongest baseline by approximately 97\%, 12\%, and 39\% on Parking, Humanoid, and DonkeyCar, respectively. To assess generalization beyond the training distribution, we construct five out-of-distribution sets spanning unseen parking slots, extrapolated vehicle poses, larger humanoid torso-height deviations, and roads with higher total curvature. \sys leads on all five, with mean \auprc\ gains of up to 145\% on Parking, 29\% on Humanoid, and 44\% on DonkeyCar. Under the same test generation budget, the best combination of \sys and a search algorithm achieves 29\%, 22\%, and 79\% higher mean novel-failure counts than the strongest baseline combination on Parking, Humanoid, and DonkeyCar, respectively. For failure diversity, varying the number of clusters $k$ from 2 to 40 shows that \sys attains the highest coverage for almost every $k$.

This paper makes the following contributions:
\begin{compactitem}
  \item \textbf{The first world-model surrogate for testing DRL agents.} We adapt an RSSM to learn the agent's behavior and environment dynamics from per-step training records that existing configuration-only surrogates discard and cannot exploit without fundamentally changing their approaches. To our knowledge, this is the first use of a world model as a surrogate for testing DRL agents.
  \item \textbf{A comprehensive evaluation of failure prediction, test generation, and failure diversity.} We evaluate failure prediction on held-out in-distribution configurations and on five out-of-distribution sets. We also evaluate search-based test generation and the diversity of the discovered failures. We identify two threats in existing evaluation protocols---stale labels recorded during training and seeded searches that replay known failures---and mitigate them by re-executing configurations with the final agent and counting only novel failures.
  \item \textbf{Evidence of effective failure prediction and test generation.} \sys achieves the highest mean prediction accuracy on all three benchmarks and all five out-of-distribution test sets. Its best combinations with the search algorithms find more novel failures on average across all three benchmarks, and its failure-diversity advantage persists across almost the entire range $k=2,\ldots,40$. Ablations examine how imagined rollouts, scoring, and training contribute to these gains.
\end{compactitem}

\section{Background}
\label{sec:background}

\subsection{Surrogate-Assisted Testing of DRL Agents}
\label{sec:background:drl}

A DRL agent interacts with an environment in \textit{episodes}: at each step $t$ it observes $o_t$, selects an action $a_t = \pi(o_t)$ with its policy network $\pi$, and receives a reward $r_t$. The episode ends when the task is completed, a failure occurs, or a time limit is reached. Each episode is instantiated from an \textit{environment configuration} $c \in \mathbb{R}^d$ drawn from a generator. The configurations on which the trained agent still fails are the targets of testing.

\indago~\cite{biagiola2024indago} formulates this problem as \textit{surrogate-assisted search}. Its input is the \textit{training log}: for every episode played while the agent was trained, the configuration $c$, the outcome $y \in \{\text{fail}, \text{pass}\}$, and, as a by-product of the training loop, the observations, actions, and rewards of every step. \indago uses only the pairs $(c, y)$ to fit a multi-layer perceptron (MLP) $f_\theta(c) \approx P(y{=}\text{fail} \mid c)$. At test time, a search algorithm maximises $f_\theta$, using a genetic algorithm (GA) or a hill climber guided by input saliency~\cite{simonyan2013saliency}, both \textit{seeded} with the recorded failures (the seeded hill climber is the recommended default). Each invocation returns one configuration, whose outcome is then validated in the simulator.  

\subsection{World Models}
\label{sec:background:wm}

A \textit{world model} learns how an environment evolves in response to an agent's actions~\cite{ha2018worldmodels}. From an initial state, it predicts subsequent observations, rewards, and episode termination, allowing episodes to be generated without interacting with the real environment. The \textit{recurrent state-space model} (RSSM)~\cite{hafner2019planet}, used in Dreamer~\cite{hafner2020dreamer,hafner2025dreamerv3}, represents this evolution through a recurrent hidden state and a stochastic latent state. During training, a \textit{posterior} infers the latent state from observed interactions, while a \textit{prior} learns to predict it without the current observation. During \textit{imagination}, the model uses the prior to generate successive latent states under a sequence of actions, without requiring new environmental observations. These models support policy learning through imagined interactions. Testing a trained agent requires a different capability: generating an episode from a candidate environment configuration and assessing whether the agent would fail. We adapt the RSSM for this purpose by learning configuration-based initialisation, modelling the trained agent's actions, and predicting termination and failure.  

\section{The \sys Surrogate}
\label{sec:method}

\subsection{Overview}
\label{sec:method:overview}

\begin{figure}[tb]
  \centering
  \includegraphics[width=\textwidth]{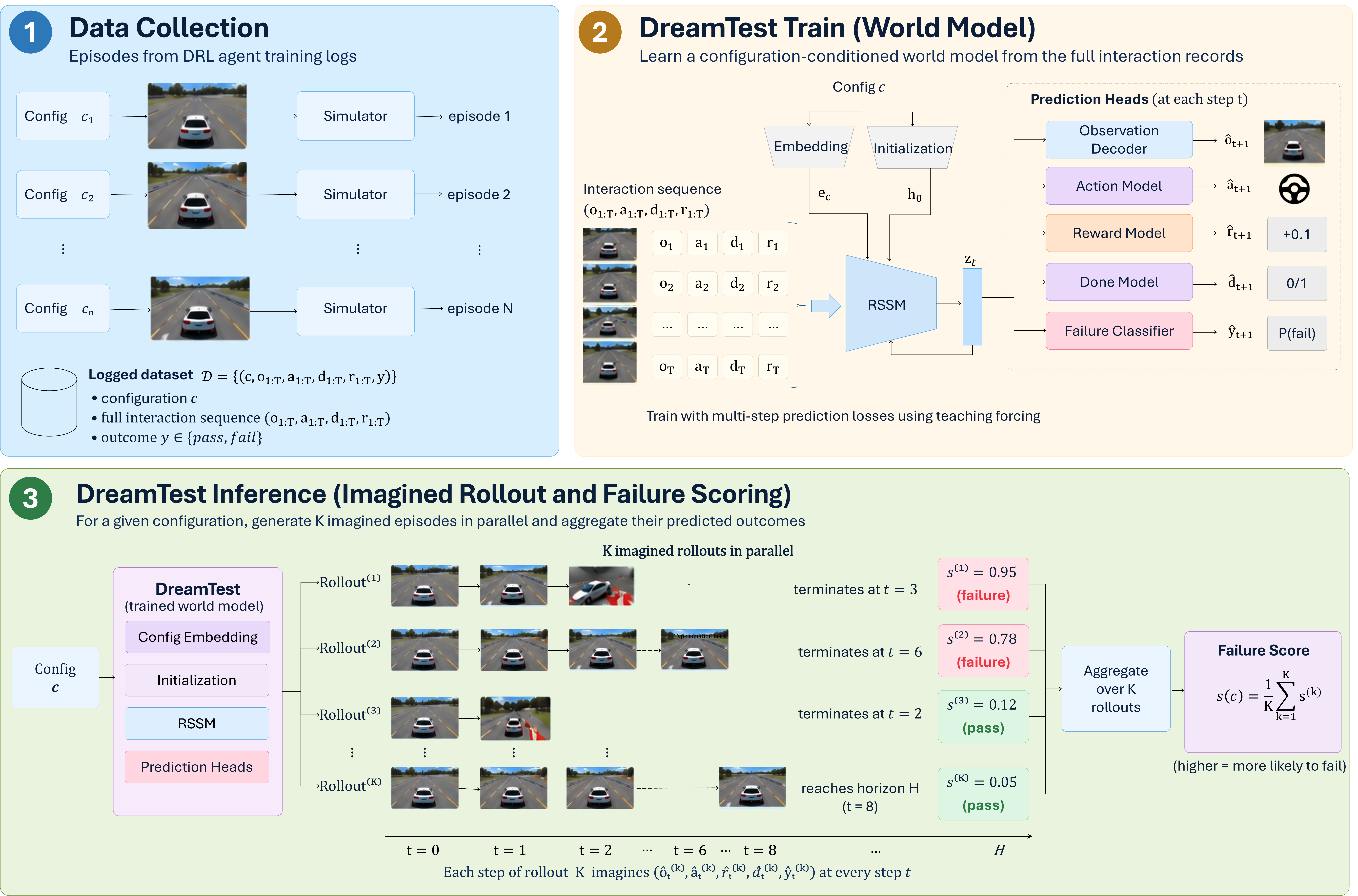}
  \caption{Workflow of \sys. }
  \label{fig:overview}
\end{figure}

\sys maps an environment configuration to a failure score by generating imagined episodes. We adapt a recurrent state-space model (RSSM) to initialise episodes from configurations, model the trained agent's actions, and predict termination and failure. \Cref{fig:overview} summarises the data collection, training, and inference stages.

\textcircled{\scriptsize 1} \textit{Data collection.} For each episode of length $T$, the DRL training log records a configuration $c \in \mathbb{R}^d$ and a trajectory $(o_{1:T}, a_{1:T}, r_{1:T}, d_{1:T})$, where $o_t$ is the observation, $a_t$ the action, $r_t$ the reward, and $d_t \in \{0,1\}$ indicates termination at step $t$ (with $d_t=1$ meaning the episode ends). The subscript $1{:}T$ denotes the full sequence.

\textcircled{\scriptsize 2} \textit{Training.} The adapted RSSM is trained to model environment dynamics and the agent's behaviour, and to predict termination and failure (\Cref{sec:method:model}).

\textcircled{\scriptsize 3} \textit{Inference.} Given a candidate configuration, \sys generates $K$ imagined episodes in parallel and aggregates their termination-weighted failure predictions to produce a score $\pfail$ (\Cref{sec:method:inference}).

\subsection{\sys Training}
\label{sec:method:model}

We first define the surrogate's inputs, output, and training data. We then describe the adapted RSSM architecture and its prediction heads, followed by the training objective and optimisation procedure.

\subsubsection{Inputs, Output, and Training Data}
\label{sec:method:data}

The surrogate takes a configuration vector $c \in \mathbb{R}^d$ and returns a failure score $\pfail \in [0,1]$. Its training dataset is $\mathcal{D} = \{(c, o_{1:T}, a_{1:T}, r_{1:T}, d_{1:T}, y)\}$, extracted from the DRL agent's training log, where each tuple corresponds to one logged episode where $y$ is its binary outcome (pass/fail). When these records are available, training the surrogate requires no additional simulator executions.

\subsubsection{Model Architecture}
\label{sec:method:transition}
\label{sec:method:heads}

Given an environment configuration $c$, the \textit{Config Embedding} block in stage \textcircled{\scriptsize 2} of \Cref{fig:overview} encodes it as a feature vector $e_c$. In parallel, the \textit{Initialisation} network $f_{\mathrm{in}}$ maps $c$ to the initial recurrent state $h_0$. This configuration-based initialisation is our first adaptation to the RSSM. Whereas the original model starts rollouts from states inferred from observations and actions~\cite{hafner2019planet}, \sys must start from a configuration that has not been executed.

Inside the \textit{Recurrent State-Space Model} block, a learned latent prediction network maps $h_0$ to a distribution $p(z_0 \mid h_0)$ over the initial latent vector. A sample $z_0$ is concatenated with $h_0$ to form the initial state feature $s_0 = [h_0, z_0]$. This stochastic sampling allows the same configuration to generate different imagined episodes.

At step $t$, the action head receives the preceding feature $s_{t-1}$ and predicts the agent's action $\hat{a}_t$, such as a steering command in a driving episode. A gated recurrent unit (GRU) computes $h_t$ deterministically from $h_{t-1}$, $z_{t-1}$, $\hat{a}_t$, and $e_c$. The recurrent state summarises the preceding interactions. Supplying $e_c$ at every transition is our second adaptation to the RSSM. For a road configuration containing a bend, this keeps the road geometry available as the imagined car approaches it.  

The updated recurrent state $h_t$ determines the distribution $p(z_t \mid h_t)$ from which the next latent vector $z_t$ is sampled. The resulting feature $s_t = [h_t, z_t]$ feeds five prediction heads, including the action head that drives the next transition. The initialisation and inference dynamics are
\begin{align}
  h_0 &= f_{\mathrm{in}}(c), \nonumber\\
  z_0 &\sim p(z_0 \mid h_0), \nonumber\\
  h_t &= \mathrm{GRU}\big(h_{t-1},\, [\,z_{t-1},\, \hat{a}_t,\, e_c,\, x_t\,]\big), \label{eq:transition}\\
  z_t &\sim p(z_t \mid h_t). \nonumber
\end{align}

The five \textit{Prediction Heads} in \Cref{fig:overview} receive $s_t$ as input. During training, the recorded observation $o_t$ is used to infer $z_t$ (\Cref{sec:method:training}). Each head predicts an observable quantity or outcome from this feature and is supervised by the corresponding logged target.

The \textit{observation decoder} predicts $\hat{o}_{t+1}$ and is trained by the mean squared error against the logged observation $o_{t+1}$, averaged over its $n_o$ components:
\begin{equation}
  \mathcal{L}_{\mathrm{obs}} = \operatorname{mean}\!\left[\frac{\|\hat{o}_{t+1} - o_{t+1}\|_2^2}{n_o}\right].
  \label{eq:loss:obs}
\end{equation}

The \textit{reward model} predicts $\hat{r}_{t+1}$ and minimises the mean squared error against the logged reward $r_{t+1}$:
\begin{equation}
  \mathcal{L}_{\mathrm{reward}} = \operatorname{mean}\!\left[(\hat{r}_{t+1} - r_{t+1})^2\right].
  \label{eq:loss:reward}
\end{equation}

The \textit{action model} learns the tested agent's behaviour through behaviour cloning. It predicts $\hat{a}_{t+1}$ from the latent feature and is supervised by the logged action $a_{t+1}$. Its mean squared error is averaged over the $n_a$ action components:
\begin{equation}
  \mathcal{L}_{\mathrm{action}} = \operatorname{mean}\!\left[\frac{\|\hat{a}_{t+1} - a_{t+1}\|_2^2}{n_a}\right].
  \label{eq:loss:action}
\end{equation}

The \textit{done model} predicts the termination probability $\hat{d}_t \in [0,1]$ and is trained by binary cross-entropy against the logged flag $d_t$:
\begin{equation}
  \mathcal{L}_{\mathrm{done}} = -\operatorname{mean}\!\left[d_t\log\hat{d}_t + (1-d_t)\log(1-\hat{d}_t)\right].
  \label{eq:loss:done}
\end{equation}

The \textit{failure classifier} predicts a failure score $\hat{y}_t$ at each step, but receives a training target only at the terminal step $T$. This target is $y_f=1$ when the logged outcome $y$ is fail and $y_f=0$ when it is pass. Its binary cross-entropy loss is
\begin{equation}
  \mathcal{L}_{\mathrm{fail}} = -\operatorname{mean}\!\left[y_f\log\hat{y}_T + (1-y_f)\log(1-\hat{y}_T)\right].
  \label{eq:loss:fail}
\end{equation}

The failure loss is averaged over terminal features only. Restricting supervision to terminal states avoids assigning a later outcome to intermediate states. During inference, the classifier is evaluated at every step, and its predictions are weighted by the termination probabilities (\Cref{eq:pfail}). \Cref{sec:results:ablation} evaluates this scoring rule.

The combined head-supervision loss is
\begin{equation}
  \mathcal{L}_{\mathrm{heads}} = \mathcal{L}_{\mathrm{obs}} + \mathcal{L}_{\mathrm{reward}} + \mathcal{L}_{\mathrm{action}} + \mathcal{L}_{\mathrm{done}} + w_f\,\mathcal{L}_{\mathrm{fail}}.
  \label{eq:loss:heads}
\end{equation}

The coefficient $w_f$ balances failure supervision against the other head objectives and is selected empirically on the validation set (\Cref{sec:design:impl}).

\subsubsection{Training Objective and Optimisation}
\label{sec:method:training}

Training uses teacher forcing on logged episodes. Given a tuple $(c, o_{1:T}, a_{1:T}, r_{1:T}, d_{1:T}, y)$, the model processes the recorded episode from beginning to end. At each step, it receives the logged observation and action, predicts the next observation, reward, and action together with termination and failure, and compares these predictions with the log (\Cref{eq:loss:heads}). Gradient descent then updates all parameters.

At inference, there is no recorded episode to follow. Given only a new configuration, the model must feed its own predictions into subsequent steps. Teacher forcing can therefore create a gap between training and inference. Following RSSM training~\cite{hafner2019planet}, we add two KL regularisers that teach the model to continue without seeing the previous logged observation. Here, $q_t$ denotes the latent distribution obtained after seeing $o_t$, while $p_t$ denotes the distribution predicted without it. The one-step loss checks this prediction after one transition. Latent overshooting repeats the exercise for $j$ transitions before comparing with the logged trajectory:
\begin{equation}
  \mathcal{L}_{\mathrm{KL}} = \operatorname{mean}\!\left[\mathrm{KL}(q_t\,\|\,p_t)\right], \qquad
  \mathcal{L}_{\mathrm{overshoot}} = \operatorname{mean}\!\left[\mathrm{KL}(q_{t+j}\,\|\,p_{t\to t+j})\right].
  \label{eq:loss:latent}
\end{equation}
Therefore, the overall training objective is
\begin{equation}
  \mathcal{L} = \mathcal{L}_{\mathrm{heads}} +  \mathcal{L}_{\mathrm{KL}} + \mathcal{L}_{\mathrm{overshoot}}.
  \label{eq:loss}
\end{equation}

\subsection{\sys Inference}
\label{sec:method:inference}

At inference, \sys requires only a candidate configuration $c$. It neither executes the simulator nor queries the trained agent's policy. Instead, it generates $K$ imagined episodes, or \textit{rollouts}, in parallel using the learned transition and action model.

Each rollout $k$ begins with $h_0 = f_{\mathrm{in}}(c)$ and $z_0 \sim p(z_0 \mid h_0)$. At step $t$, the action model predicts $\hat{a}_t = \mathrm{action}(s_{t-1})$. The transition in \Cref{eq:transition} advances the recurrent state, and the prior generates the next stochastic state. The outcome heads then produce termination and failure predictions $\hat{d}_{k,t}$ and $\hat{y}_{k,t}$. Each rollout retains a fixed dropout mask~\cite{gal2016dropout}, representing one coherent sample of the model throughout the episode.

A rollout stops when its predicted survival probability $S_{k,t} = \prod_{s \le t}(1 - \hat{d}_{k,s})$ falls below a threshold $\epsilon$ or reaches the horizon $H$. We set $S_{k,0}=1$. The probability of termination at step $t$ is $\mathrm{term}_{k,t} = S_{k,t-1}\,\hat{d}_{k,t}$, which combines survival up to that step with termination there. Weighting the failure prediction by this probability accounts for uncertainty about when the episode ends. The resulting rollout scores are averaged to obtain the configuration's score:
\begin{equation}
  s^{(k)} = \sum_{t} \mathrm{term}_{k,t}\;\hat{y}_{k,t}, \qquad
  \pfail = \frac{1}{K}\sum_{k=1}^{K} s^{(k)}.
  \label{eq:pfail}
\end{equation}
When a rollout stops, we assign its remaining survival probability to the final visited step so that the termination weights sum to one. For a rollout that reaches $H$, this assigns the probability of continuing beyond the horizon to the failure prediction at the last imagined state.  
 
\section{Experimental Design}
\label{sec:design}

\subsection{Research Questions}
\label{sec:design:rqs}

We study three research questions (RQs):

\begin{compactitem}
  \item \textbf{RQ1: Effectiveness of failure prediction.} How effectively does \sys predict failures, compared with existing surrogates, on in-distribution and out-of-distribution configurations?
  \item \textbf{RQ2: Effectiveness of test generation.} Under a fixed simulator-validation budget, how effectively does \sys-guided test generation discover novel failures, and how does it vary across search algorithms?
  \item \textbf{RQ3: Failure diversity.} Under a fixed simulator-validation budget, how many distinct failure behaviours does \sys discover, compared with the baselines?
\end{compactitem}

RQ1 isolates the quality of the surrogate itself: effective search guidance requires failures to be ranked above passes, both in-distribution and under distribution shift. We therefore evaluate single models trained on ten data splits on held-out in-distribution and out-of-distribution configurations. RQ2 then examines whether predictive performance translates into practical test generation, since the search algorithm may affect which failures are ultimately found. Under a common validation budget, we use each surrogate to guide the same ranking, genetic, and hill-climbing algorithms and count the novel failures they discover. Because a single fitted model can be uncertain and less robust, surrogates are often ensembled~\cite{goel2007ensemble,lakshminarayanan2017ensembles,chua2018pets}; we therefore report both single-model results and ensembles that average ten independently trained models of the same type. Finally, RQ3 measures how many distinct failure behaviours each technique discovers under the same validation budget. We cluster execution traces and compare the number of behavioural clusters covered by each technique. Thus, RQ2 excludes exact replays of known failures, while RQ3 checks whether the remaining failures are behaviourally diverse rather than small perturbations that reproduce the same failure behaviour.

\subsection{Benchmarks and Datasets}
\label{sec:design:subjects}

The published benchmarks provide episode logs collected during DRL training. Each logged episode contains a test configuration and the pass/fail outcome produced by the agent at that point in training. Because the agent may improve during later training, these logs can exhibit label staleness: a configuration recorded as failing may pass when re-executed with the final trained agent. We refer to the recorded outcomes as \textit{historical} labels. Prior evaluations of the baseline surrogates treat these labels as ground truth without re-executing configurations. To address this issue, we re-execute held-out configurations with the final agent and refer to the resulting outcomes as \textit{fresh} labels. For in-distribution sets, we report results using both label types. We also evaluate separately generated out-of-distribution sets labelled directly with the final agent.

\subsubsection{In-Distribution Data}
\label{sec:design:id-data}

We evaluate on three benchmarks from the DRL-testing literature~\cite{biagiola2024indago,bartlett2025nexus}. They span diverse application domains, DRL algorithms, and configuration spaces: autonomous parking, robotic locomotion, and autonomous driving.

\noindent\textit{Parking.} In HighwayEnv~\cite{leurent2018highway}, an agent trained with hindsight experience replay (HER)~\cite{andrychowicz2017her} must reach a goal slot while avoiding parked vehicles. The configuration encodes the goal slot, ego pose, and slot occupancy ($d = 24$). An episode fails if the vehicle collides or does not park. The dataset contains 5\,099 episodes, including 132 failures (2.6\%). Among the 4\,547 configurations for which we obtained fresh labels, 91 (2.0\%) changed labels (71 failure$\to$pass; 20 pass$\to$failure), reducing the failure rate from 2.6\% to 1.4\%.

\noindent\textit{Humanoid.} In MuJoCo~\cite{todorov2012mujoco}, an agent trained with truncated quantile critics (TQC)~\cite{kuznetsov2020tqc} must keep a humanoid upright for 300 simulator steps. The configuration is the initial 47-dimensional joint state, perturbed within a box of half-width $0.03$ around the reference pose ($d = 47$). An episode fails when the humanoid falls. The dataset contains 12\,000 episodes, including 836 failures (7.0\%). Among the 10\,658 configurations with fresh labels, 845 (7.9\%) changed labels (439 failure$\to$pass; 406 pass$\to$failure), reducing the failure rate from 7.0\% to 6.6\%.

\noindent\textit{DonkeyCar.} In the Unity simulator~\cite{donkeysim}, an agent trained with soft actor--critic (SAC)~\cite{haarnoja2018sac} drives from camera images around a procedurally generated track. The configuration specifies twelve track elements, each represented by a command and parameter ($d = 24$). An episode fails when the vehicle leaves the road. The dataset contains 5\,779 episodes, including 663 failures (11.5\%). Among the 5\,162 configurations with fresh labels, 518 (10.0\%) changed labels (227 failure$\to$pass; 291 pass$\to$failure), increasing the failure rate from 11.5\% to 12.8\%.

\subsubsection{Out-of-Distribution Data}
\label{sec:design:ood-data}

We evaluate two Parking out-of-distribution sets: configurations with unseen goal slots, and independently generated configurations with extrapolated starting positions. For Humanoid, we draw a random configuration pool and select the top 20\% and top 5\% by initial torso-height deviation. For DonkeyCar, we select the top 20\% by summed road curvature. We execute their respective DRL agents on their simulators to label every configuration. For each set, the ten repetitions train on the same in-distribution data and differ only in training seed.

\subsection{Baselines}
\label{sec:design:baselines}

We use separate baseline sets for failure prediction and test generation.

\noindent\textit{Failure-prediction baselines (RQ1).} We compare against five configuration-based surrogates used in prior DRL-testing work. From \indago~\cite{biagiola2024indago}, \textit{MLP} is a binary classifier over configurations, while \textit{Regressor} predicts a continuous measure of failure severity. From \samota~\cite{haq2022samota}, \textit{Kriging} fits a probabilistic response surface, the \textit{Polynomial} model uses polynomial features, and the \textit{RBF network} uses radial basis functions to model the configuration--outcome relationship.

\noindent\textit{Test-generation baselines (RQ2 and RQ3).} Surrogate-assisted test generation combines a surrogate model with a search algorithm to propose configurations for validation. Because our focus is the surrogate rather than search design, we adopt the ranking, genetic, and hill-climbing algorithms from \indago~\cite{biagiola2024indago}. We pair \sys and each surrogate baseline with every compatible algorithm to obtain a controlled downstream comparison.

\textit{Ranking} samples a candidate pool, orders it by surrogate score, and validates the highest-ranked configurations~\cite{uesato2019rigorous}. The seeded \textit{genetic algorithm} (GA) mutates recorded failures and evolves them towards higher surrogate scores. The seeded \textit{saliency hill climber} (HC) updates a recorded failure using the gradient of the surrogate score and therefore applies only to \sys and the MLP.

We also compare against end-to-end test generation baselines. \nexus~\cite{bartlett2025nexus} is a many-objective testing technique that uses \indago's MLP internally. Its released implementation supports only Parking, so we evaluate it only on that benchmark. \mdpfuzz~\cite{pang2022mdpfuzz} is a model-free fuzzer for agents solving Markov decision processes (MDPs) that executes every candidate in the simulator. We use the re-implementation from its replicability study~\cite{mazouni2024mdpfuzzrep}, with Gaussian mixture model (GMM) coverage guidance (\mdpfuzz) and without it (\mdpfuzz$^{-}$). \textit{Random} sample configurations from the environment generator. \textit{Training-failure replay} re-executes recorded failures and therefore measures rediscovery rather than discovery.

\subsection{Evaluation Metrics}
\label{sec:design:metrics}

\noindent\textit{RQ1: Failure prediction.} Each surrogate assigns a failure score to every configuration. Given a decision threshold $\tau$, configurations with score $\ge \tau$ are predicted to fail. From the resulting true positives (TP), false positives (FP), true negatives (TN), and false negatives (FN), we compute precision, recall (true-positive rate; TPR), and the false-positive rate (FPR):
\begin{equation}
  \mathrm{precision}(\tau) = \frac{\mathrm{TP}}{\mathrm{TP}+\mathrm{FP}}, \qquad
  \mathrm{recall}(\tau) = \mathrm{TPR}(\tau) = \frac{\mathrm{TP}}{\mathrm{TP}+\mathrm{FN}}, \qquad
  \mathrm{FPR}(\tau) = \frac{\mathrm{FP}}{\mathrm{FP}+\mathrm{TN}}.
  \label{eq:metrics:rates}
\end{equation}
Sweeping $\tau$ yields a precision--recall (PR) curve and a receiver operating characteristic (ROC) curve. We report the area under the PR curve (\auprc) and the area under the ROC curve (\auroc)~\cite{davis2006prroc,saito2015prc}. We use \auprc\ as the primary metric because failures are rare: recall measures how many failures are found, while precision penalises configurations incorrectly ranked as failures. Under random ranking, \auprc\ equals the failure rate, which we report as the base rate. We use \auroc\ as a secondary measure of score separation between failures and passes; $0.5$ denotes random ranking and $1$ perfect separation. Unlike precision, recall, or the F1 score at a single threshold, both area measures summarise performance across all thresholds and avoid choosing an arbitrary decision threshold.

\noindent\textit{RQ2: Test generation.} Each surrogate-guided method generates test configurations without querying the simulator. For evaluation, we execute the generated tests in the simulator with the trained agent to obtain ground-truth pass/fail outcomes.

A limitation of prior evaluations is that they count every observed failure, even though seeded search algorithms start from failures in the training log and may simply regenerate known cases. Raw failure counts therefore conflate failure discovery with rediscovery. We instead use the number of \textit{novel failures} as the primary metric. A generated test counts as a novel failure if it fails during validation and its configuration does not match any configuration recorded as failing in the training log. We consider two configurations to match if every coordinate is equal after rounding to four decimal places.

\noindent\textit{RQ3: Failure diversity.} RQ2 excludes exact replays of failures in the training log. RQ3 goes further by checking whether the remaining configuration-novel failures exhibit diverse behaviours rather than being small perturbations around known failures that reproduce the same behaviour. Such diversity is important because different failure behaviours may expose different defects. We therefore cluster the execution traces of the configuration-novel failures found in each run. Following \indago~\cite{biagiola2024indago}, we represent each failing execution as a time series sampled at every simulator step. For Parking, the series records the ego vehicle's two-dimensional position (its trajectory). For Humanoid, it records torso height, capturing the dynamics of the fall. For DonkeyCar, it records the vehicle's cross-track error, which is its lateral distance from the road centreline and captures how it leaves the road.

Because executions can differ in length, we pad shorter time series with zeros to a common length and flatten each series into a fixed-length vector. We then apply $k$-means to group failures with similar behaviours. A run covers a cluster if it finds at least one configuration-novel failure assigned to that cluster. Let $g_k(e)$ denote the cluster index assigned to failure $e$ when clustering into $k$ clusters. For the set $E$ of configuration-novel failures found in a run, cluster coverage is the number of clusters represented in $E$:
\begin{equation}
  N_k(E) = \left|\{g_k(e) : e \in E\}\right|,
  \qquad k \in \{2,3,\ldots,40\}.
  \label{eq:metrics:diversity}
\end{equation}
We emphasise that \sys and every baseline receive the same task and budget: each method submits 50 configurations for simulator validation, which can poteintially cause failures according to their predictions to ensure a fare comparison. 
\indago selects the number of clusters $k$ using the silhouette coefficient~\cite{biagiola2024indago,rousseeuw1987silhouette}. However, reporting coverage at a single value of $k$ can bias the comparison because the result depends on how finely failures are partitioned. We therefore sweep $k$ from 2 to 40. For each $k$, we use the same cluster definitions for every technique and report the mean $N_k(E)$ over the ten runs.

\subsection{Statistical Testing}
\label{sec:design:statistics}

We repeat every experiment ten times. For each comparison, we pair the results for \sys and the baseline by data split or random seed and apply the two-sided Wilcoxon signed-rank test. We use $p<0.05$ as the significance threshold~\cite{vargha2000a12,arcuri2014hitchhiker}.

\subsection{Implementation and Setup}
\label{sec:design:impl}

\noindent\textit{Training and hyperparameter tuning.} Following \indago, we create ten stratified 80/20 training--test splits and reserve a stratified 15\% subset of each training partition for validation. All surrogate models use the same dataset and splits, with a shared budget of 60 hyperparameter trials. \Cref{tab:design:hparams} reports the selected hyperparameters for \sys and the five baseline surrogates, omitting fixed architecture and inference settings.

\begin{table}[tb]
  \centering
  \caption{Hyperparameters selected for \sys and baselines}
  \label{tab:design:hparams}
  \scriptsize
  \setlength{\tabcolsep}{5pt}
  \resizebox{\textwidth}{!}{%
  \begin{tabular}{@{}>{\centering\arraybackslash}m{1.2cm}*{6}{>{\centering\arraybackslash}m{2.55cm}}@{}}
    \toprule
    Benchmark & \sys & MLP & Regressor & Polynomial & Kriging & RBF network \\
    \midrule
    Parking
      & \shortstack[c]{\texttt{overshoot}=0 \\ \texttt{ov\_starts}=8 \\ $\mathrm{lr}=6.64\times10^{-4}$ \\ $\beta=1.91$ \\ free bits${}=1$}
      & \shortstack[c]{$4\times32$ hidden units \\ $\mathrm{lr}=3\times10^{-4}$ \\ epochs${}=200$ \\ \texttt{fail\_frac}${}=0.25$ \\ batch${}=256$}
      & \shortstack[c]{hidden${}=(64,64)$ \\ $\alpha=10^{-3}$ \\ $\mathrm{lr}=10^{-3}$ \\ \texttt{max\_iter}${}=600$ \\ patience${}=20$}
      & \shortstack[c]{degree${}=2$ \\ $C=0.0012$ \\ \texttt{class\_weight}=none}
      & \shortstack[c]{$\ell=1.92$ \\ constant${}=0.20$ \\ subsample${}=3\,000$}
      & \shortstack[c]{\texttt{n\_centers}${}=25$ \\ $\gamma=0.080$ \\ $C=55$} \\
    \midrule
    Humanoid
      & \shortstack[c]{\texttt{overshoot}=12 \\ \texttt{ov\_starts}=16 \\ $\mathrm{lr}=7.62\times10^{-4}$ \\ $\beta=0.97$ \\ free bits${}=2$}
      & \shortstack[c]{$4\times32$ hidden units \\ $\mathrm{lr}=3\times10^{-4}$ \\ epochs${}=200$ \\ \texttt{fail\_frac}${}=0.25$ \\ batch${}=256$}
      & \shortstack[c]{hidden${}=(64,64)$ \\ $\alpha=10^{-3}$ \\ $\mathrm{lr}=10^{-3}$ \\ \texttt{max\_iter}${}=600$ \\ patience${}=20$}
      & \shortstack[c]{degree${}=2$ \\ $C=0.0012$ \\ \texttt{class\_weight}=none}
      & \shortstack[c]{$\ell=12.1$ \\ constant${}=0.13$ \\ subsample${}=3\,000$}
      & \shortstack[c]{\texttt{n\_centers}${}=200$ \\ $\gamma=0.031$ \\ $C=36$} \\
    \midrule
    DonkeyCar
      & \shortstack[c]{\texttt{overshoot}=5 \\ \texttt{ov\_starts}=8 \\ $\mathrm{lr}=6.79\times10^{-4}$ \\ $\beta=1.57$ \\ free bits${}=0.25$}
      & \shortstack[c]{$4\times32$ hidden units \\ $\mathrm{lr}=3\times10^{-4}$ \\ epochs${}=200$ \\ \texttt{fail\_frac}${}=0.25$ \\ batch${}=256$}
      & \shortstack[c]{hidden${}=(64,64)$ \\ $\alpha=10^{-3}$ \\ $\mathrm{lr}=10^{-3}$ \\ \texttt{max\_iter}${}=600$ \\ patience${}=20$}
      & \shortstack[c]{degree${}=3$ \\ $C=0.11$ \\ \texttt{class\_weight}=balanced}
      & \shortstack[c]{$\ell=0.65$ \\ constant${}=7.42$ \\ subsample${}=2\,000$}
      & \shortstack[c]{\texttt{n\_centers}${}=200$ \\ $\gamma=0.0065$ \\ $C=55$} \\
    \bottomrule
  \end{tabular}
  }
\end{table}

\noindent\textit{Search budget.} We run every technique ten times. For each surrogate-guided method, the search stops after producing 50 configurations predicted to fail; candidate generation queries only the surrogate, and we then validate each of these 50 configurations once in the simulator. Ranking considers a pool of 3\,000 candidates, while the GA and hill climber receive at most 5\,000 surrogate queries per proposal. We apply these budgets to every compatible surrogate--search combination and evaluate both single models and ten-model ensembles.  Experiments ran on a workstation equipped with two NVIDIA RTX GPUs and 64 CPU cores.

\section{Results}
\label{sec:results}

\subsection{RQ1: Effectiveness of Failure Prediction}
\label{sec:results:rq1}

\subsubsection{In-distribution configurations}
\label{sec:results:rq1:id}

\begin{table}[tb]
  \centering
  \caption{RQ1, failure-prediction effectiveness on in-distribution configurations, reported as mean $\pm$ standard deviation over ten data splits. \textit{Hist.}: labels recorded in the training log. \textit{Fresh}: labels obtained by re-executing the same configurations with the final agent. $^{*}$: significantly different from \sys under the two-sided Wilcoxon signed-rank test ($p<0.05$).}
  \label{tab:rq1:id}
  \scriptsize
  \renewcommand{\arraystretch}{0.95}
  \setlength{\tabcolsep}{3.5pt}
  \begin{tabular}{@{}llcccc@{}}
    \toprule
    & & \multicolumn{2}{c}{\auprc} & \multicolumn{2}{c}{\auroc} \\
    \cmidrule(lr){3-4}\cmidrule(lr){5-6}
    Benchmark & Surrogate & hist. & fresh & hist. & fresh \\
    \midrule
    \multirow{6}{*}{Parking}
      & \sys & \textbf{0.156$\pm$0.035} & \textbf{0.120$\pm$0.063} & \textbf{0.832$\pm$0.040} & \textbf{0.817$\pm$0.071} \\
      & MLP & 0.074$\pm$0.028$^{*}$ & 0.061$\pm$0.041$^{*}$ & 0.648$\pm$0.050$^{*}$ & 0.684$\pm$0.116$^{*}$ \\
      & Regressor & 0.048$\pm$0.025$^{*}$ & 0.030$\pm$0.034$^{*}$ & 0.567$\pm$0.060$^{*}$ & 0.527$\pm$0.076$^{*}$ \\
      & Kriging & 0.058$\pm$0.018$^{*}$ & 0.048$\pm$0.037$^{*}$ & 0.703$\pm$0.034$^{*}$ & 0.674$\pm$0.072$^{*}$ \\
      & Polynomial & 0.058$\pm$0.018$^{*}$ & 0.049$\pm$0.039$^{*}$ & 0.708$\pm$0.029$^{*}$ & 0.683$\pm$0.069$^{*}$ \\
      & RBF network & 0.054$\pm$0.010$^{*}$ & 0.039$\pm$0.018$^{*}$ & 0.720$\pm$0.033$^{*}$ & 0.708$\pm$0.076$^{*}$ \\
    \midrule
    \multirow{6}{*}{Humanoid}
      & \sys & \textbf{0.210$\pm$0.030} & \textbf{0.198$\pm$0.045} & \textbf{0.806$\pm$0.025} & \textbf{0.802$\pm$0.021} \\
      & MLP & 0.153$\pm$0.021$^{*}$ & 0.149$\pm$0.031$^{*}$ & 0.705$\pm$0.029$^{*}$ & 0.710$\pm$0.030$^{*}$ \\
      & Regressor & 0.136$\pm$0.012$^{*}$ & 0.132$\pm$0.019$^{*}$ & 0.685$\pm$0.013$^{*}$ & 0.681$\pm$0.019$^{*}$ \\
      & Kriging & 0.152$\pm$0.013$^{*}$ & 0.148$\pm$0.013$^{*}$ & 0.736$\pm$0.015$^{*}$ & 0.733$\pm$0.010$^{*}$ \\
      & Polynomial & 0.178$\pm$0.020 & 0.177$\pm$0.023 & 0.756$\pm$0.014$^{*}$ & 0.762$\pm$0.012$^{*}$ \\
      & RBF network & 0.170$\pm$0.012$^{*}$ & 0.167$\pm$0.015 & 0.753$\pm$0.010$^{*}$ & 0.751$\pm$0.015$^{*}$ \\
    \midrule
    \multirow{6}{*}{DonkeyCar}
      & \sys & \textbf{0.393$\pm$0.037} & \textbf{0.422$\pm$0.062} & \textbf{0.830$\pm$0.029} & \textbf{0.829$\pm$0.029} \\
      & MLP & 0.213$\pm$0.023$^{*}$ & 0.235$\pm$0.029$^{*}$ & 0.696$\pm$0.027$^{*}$ & 0.690$\pm$0.022$^{*}$ \\
      & Regressor & 0.197$\pm$0.022$^{*}$ & 0.204$\pm$0.022$^{*}$ & 0.678$\pm$0.029$^{*}$ & 0.676$\pm$0.028$^{*}$ \\
      & Kriging & 0.242$\pm$0.017$^{*}$ & 0.247$\pm$0.020$^{*}$ & 0.723$\pm$0.015$^{*}$ & 0.715$\pm$0.017$^{*}$ \\
      & Polynomial & 0.263$\pm$0.024$^{*}$ & 0.304$\pm$0.030$^{*}$ & 0.722$\pm$0.021$^{*}$ & 0.732$\pm$0.008$^{*}$ \\
      & RBF network & 0.239$\pm$0.016$^{*}$ & 0.236$\pm$0.031$^{*}$ & 0.720$\pm$0.016$^{*}$ & 0.702$\pm$0.019$^{*}$ \\
    \bottomrule
  \end{tabular}
\end{table}
~\\
\textbf{Historical labels.} \sys achieves the highest mean \auprc\ and \auroc\ in all three environments (\Cref{tab:rq1:id}). Its advantage is significant in every comparison except \auprc\ against the Polynomial model on Humanoid, where \sys still attains the higher mean (0.210 vs.\ 0.178) but the difference is not significant at $\alpha=0.05$ ($p = 0.074$). Overall, \sys is either significantly better than or statistically comparable to the strongest baseline in every case.

\textbf{Fresh labels.} Re-labelling test configurations with the final agent changes 2.1\%, 7.8\%, and 10.1\% of the labels (\Cref{sec:discussion:threats}) and lowers the Parking base rate to 1.4\%. \sys retains the highest mean \auprc\ and \auroc\ in every environment. Its advantage is significant in every comparison except \auprc\ against the Polynomial and RBF network baselines on Humanoid. In these exceptions, \sys still achieves the highest mean and is statistically comparable to the strongest baseline.

\subsubsection{Out-of-distribution configurations}
\label{sec:results:rq1:ood}

\begin{table}[tb]
  \centering
  \caption{RQ1, failure-prediction effectiveness on five out-of-distribution sets, reported as mean \auprc\ $\pm$ standard deviation and mean \auroc\ over ten training seeds. $^{*}$: significantly different from \sys in \auprc\ under the two-sided Wilcoxon signed-rank test ($p<0.05$).}
  \label{tab:rq1:ood}
  \scriptsize
  \setlength{\tabcolsep}{3pt}
  \resizebox{\textwidth}{!}{%
  \begin{tabular}{@{}l*{10}{c}@{}}
    \toprule
    & \multicolumn{5}{c}{\auprc} & \multicolumn{5}{c}{\auroc} \\
    \cmidrule(lr){2-6}\cmidrule(lr){7-11}
    Surrogate
      & unseen slots & extrapolated poses & torso 20\% & torso 5\% & severe 20\%
      & unseen slots & extrapolated poses & torso 20\% & torso 5\% & severe 20\% \\
    \midrule
    \sys
      & \textbf{0.067$\pm$0.016} & \textbf{0.203$\pm$0.037} & \textbf{0.298$\pm$0.046} & \textbf{0.327$\pm$0.031} & \textbf{0.475$\pm$0.050}
      & 0.63 & \textbf{0.81} & \textbf{0.78} & \textbf{0.77} & \textbf{0.75} \\
    MLP
      & 0.039$\pm$0.012$^{*}$ & 0.066$\pm$0.010$^{*}$ & 0.221$\pm$0.015$^{*}$ & 0.230$\pm$0.026$^{*}$ & 0.299$\pm$0.015$^{*}$
      & 0.43 & 0.58 & 0.67 & 0.64 & 0.60 \\
    Regressor
      & 0.045$\pm$0.006$^{*}$ & 0.061$\pm$0.013$^{*}$ & 0.207$\pm$0.012$^{*}$ & 0.246$\pm$0.023$^{*}$ & 0.263$\pm$0.010$^{*}$
      & 0.51 & 0.54 & 0.65 & 0.67 & 0.57 \\
    Kriging
      & 0.051$\pm$0.008$^{*}$ & 0.079$\pm$0.012$^{*}$ & 0.182$\pm$0.008$^{*}$ & 0.207$\pm$0.011$^{*}$ & 0.287$\pm$0.009$^{*}$
      & 0.60 & 0.64 & 0.64 & 0.62 & 0.59 \\
    Polynomial
      & 0.047$\pm$0.000$^{*}$ & 0.083$\pm$0.000$^{*}$ & 0.231$\pm$0.000$^{*}$ & 0.303$\pm$0.000 & 0.329$\pm$0.000$^{*}$
      & 0.59 & 0.67 & 0.69 & 0.69 & 0.63 \\
    RBF network
      & 0.058$\pm$0.008 & 0.067$\pm$0.006$^{*}$ & 0.205$\pm$0.004$^{*}$ & 0.251$\pm$0.012$^{*}$ & 0.272$\pm$0.001$^{*}$
      & \textbf{0.64} & 0.64 & 0.67 & 0.67 & --- \\
    \bottomrule
  \end{tabular}}
\end{table}

On the five out-of-distribution sets in \Cref{tab:rq1:ood}, \sys achieves the highest mean \auprc\ throughout. Its advantage is significant in 23 of the 25 baseline comparisons; the two exceptions are the strongest baselines on Unseen slots and the Humanoid torso-5\% tail, where \sys retains the higher mean but the difference is not significant at $\alpha=0.05$. \sys also achieves the highest reported \auroc\ on four sets. On Unseen slots, the RBF network is only 0.01 higher (0.64 vs.\ 0.63), with no significant difference. Overall, \sys is either the best method or statistically comparable to the best on every reported out-of-distribution comparison.

\rqconclusion{RQ1}{\sys achieves the highest mean \auprc\ and \auroc\ in every in-distribution setting. On out-of-distribution configurations, it achieves the highest mean \auprc\ on all five sets and the highest reported \auroc\ on four. In the few comparisons where it is neither best nor significantly better, \sys remains statistically comparable to the strongest baseline at $\alpha=0.05$. Overall, \sys provides more effective failure prediction than the configuration-only surrogates, including under distribution shift.}

\subsection{RQ2: Effectiveness of Test Generation}
\label{sec:results:rq2}

\begin{table}[tb]
  \centering
  \caption{RQ2 test-generation effectiveness, measured as the number of novel failures confirmed by the simulator among 50 configurations proposed by each method. Values are reported as mean $\pm$ standard deviation over ten runs. For surrogate methods, Rank, GA, and HC denote ranking, the genetic algorithm, and hill climbing. \textit{Type} distinguishes ten-model ensembles, single surrogates, and non-surrogate techniques. Bold marks the highest mean within each type and column. $^{*}$ marks a significant difference from the corresponding \sys result in the same type and column under the two-sided Wilcoxon signed-rank test ($p<0.05$). A dash indicates that the combination is not applicable.}
  \label{tab:rq2:generation}
  \footnotesize
  \setlength{\tabcolsep}{2.5pt}
  \resizebox{0.8\textwidth}{!}{%
  \begin{tabular}{@{}ll*{9}{c}@{}}
    \toprule
    & & \multicolumn{3}{c}{Parking} & \multicolumn{3}{c}{Humanoid} & \multicolumn{3}{c}{DonkeyCar} \\
    \cmidrule(lr){3-5}\cmidrule(lr){6-8}\cmidrule(lr){9-11}
    Type & Method & Rank & GA & HC & Rank & GA & HC & Rank & GA & HC \\
    \midrule
    \multirow{6}{*}{\rotatebox[origin=c]{90}{\shortstack{Ensembled\\surrogate}}} & \sys & \textbf{6.6$\pm$1.8} & 10.1$\pm$3.2 & \textbf{14.5$\pm$2.9} & \textbf{21.5$\pm$3.5} & \textbf{21.4$\pm$3.5} & \textbf{19.9$\pm$2.4} & \textbf{12.2$\pm$3.0} & 3.8$\pm$3.5 & \textbf{4.7$\pm$1.7} \\
     & MLP & 2.7$\pm$1.8$^{*}$ & \textbf{11.2$\pm$2.8} & 5.7$\pm$1.6$^{*}$ & 14.3$\pm$2.3$^{*}$ & 13.4$\pm$2.4$^{*}$ & 17.6$\pm$3.8 & 6.4$\pm$2.1$^{*}$ & 5.5$\pm$1.6 & 4.6$\pm$2.0 \\
     & Regressor & 1.0$\pm$0.9$^{*}$ & 2.6$\pm$1.5$^{*}$ & --- & 14.5$\pm$3.0$^{*}$ & 13.7$\pm$2.5$^{*}$ & --- & 4.3$\pm$2.4$^{*}$ & \textbf{6.7$\pm$2.3} & --- \\
     & Kriging & 1.3$\pm$0.8$^{*}$ & 1.4$\pm$1.3$^{*}$ & --- & 7.6$\pm$2.5$^{*}$ & 3.1$\pm$2.4$^{*}$ & --- & 6.7$\pm$1.3$^{*}$ & 1.5$\pm$1.0 & --- \\
     & Polynomial & 1.6$\pm$1.0$^{*}$ & 1.5$\pm$1.2$^{*}$ & --- & 13.1$\pm$3.2$^{*}$ & 10.1$\pm$2.3$^{*}$ & --- & 5.4$\pm$2.8$^{*}$ & 1.6$\pm$1.8 & --- \\
     & RBF network & 1.6$\pm$1.3$^{*}$ & 0.0$\pm$0.0$^{*}$ & --- & 11.9$\pm$1.7$^{*}$ & 6.0$\pm$1.6$^{*}$ & --- & 6.8$\pm$3.1$^{*}$ & 0.9$\pm$1.0 & --- \\
    \midrule
    \multirow{6}{*}{\rotatebox[origin=c]{90}{Single surrogate}} & \sys & \textbf{4.6$\pm$1.3} & \textbf{9.7$\pm$5.1} & \textbf{10.2$\pm$3.2} & \textbf{12.7$\pm$3.0} & 8.9$\pm$5.5 & 10.7$\pm$3.0 & \textbf{7.7$\pm$3.8} & \textbf{5.6$\pm$6.2} & 5.3$\pm$2.0 \\
     & MLP & 2.0$\pm$1.8$^{*}$ & 5.4$\pm$2.1$^{*}$ & 4.8$\pm$2.4$^{*}$ & 11.6$\pm$2.0 & 9.4$\pm$2.7 & \textbf{10.8$\pm$3.3} & 4.2$\pm$2.1 & 0.4$\pm$0.7$^{*}$ & \textbf{5.7$\pm$1.2} \\
     & Regressor & 1.2$\pm$1.1$^{*}$ & 3.8$\pm$3.8$^{*}$ & --- & 9.3$\pm$2.6 & 9.6$\pm$3.1 & --- & 3.8$\pm$1.7$^{*}$ & 5.1$\pm$3.9$^{n{=}7}$ & --- \\
     & Kriging & 1.2$\pm$0.9$^{*}$ & 2.8$\pm$2.1$^{*}$ & --- & 8.4$\pm$3.7 & 3.8$\pm$2.3 & --- & 5.5$\pm$1.3 & 3.1$\pm$2.3 & --- \\
     & Polynomial & 1.6$\pm$1.4$^{*}$ & 1.4$\pm$1.3$^{*}$ & --- & 11.7$\pm$2.5 & \textbf{10.0$\pm$2.5} & --- & 6.3$\pm$1.9 & 2.9$\pm$1.9 & --- \\
     & RBF network & 1.9$\pm$1.7$^{*}$ & 0.3$\pm$0.7$^{*}$ & --- & 9.9$\pm$1.7 & 6.5$\pm$1.9 & --- & 6.1$\pm$2.3 & 0.9$\pm$1.0 & --- \\
    \midrule
    \multirow{3}{*}{\rotatebox[origin=c]{90}{\scriptsize\shortstack{Non-\\surrogate}}} & \nexus & \multicolumn{3}{c}{5.7$\pm$2.2} & \multicolumn{3}{c}{---} & \multicolumn{3}{c}{---} \\
     & \mdpfuzz\ / \mdpfuzz$^{-}$ & \multicolumn{3}{c}{0.5$\pm$0.7 / 0.7$\pm$0.8} & \multicolumn{3}{c}{3.8$\pm$2.4 / 3.6$\pm$1.7} & \multicolumn{3}{c}{0.7$\pm$1.5 / 0.4$\pm$0.5} \\
     & Random / replay & \multicolumn{3}{c}{0.5$\pm$0.8 / 0.0$\pm$0.0} & \multicolumn{3}{c}{4.0$\pm$2.3 / 0.0$\pm$0.0} & \multicolumn{3}{c}{0.3$\pm$0.7 / 0.0$\pm$0.0} \\
    \bottomrule
  \end{tabular}}
\end{table}

\Cref{tab:rq2:generation} compares three categories of test-generation methods. The first couples a ten-model surrogate ensemble with a search algorithm; the second uses a single surrogate model with the same search procedure; and the third comprises fully non-surrogate approaches. Each cell reports the number of novel failures confirmed among the 50 proposed configurations.

\textbf{Ensembled surrogate and search.} The top-performing method in every environment uses the ensembled \sys. With \sys, hill climbing finds 14.5 novel failures on parking, and ranking finds 21.5 on the humanoid and 12.2 on DonkeyCar. The best competing ensemble--search combinations achieve 11.2, 17.6, and 6.8, yielding gains of 29\%, 22\%, and 79\%. These improvements are significant on parking ($p = 0.019$) and DonkeyCar ($p = 0.005$), but not on Humanoid ($p = 0.10$). Under ranking alone, the ensembled \sys also exceeds every ensembled baseline in all three environments. The weaker DonkeyCar results under the seeded algorithms arise because they spend much of the validation budget replaying known failures, whereas ranking evaluates a fresh candidate pool.

\textbf{Single surrogate and search.} A single \sys achieves the highest mean among all single-surrogate methods in every environment. Its best search combination finds 10.2 novel failures on parking, 12.7 on the humanoid, and 7.7 on DonkeyCar, compared with 5.4, 10.8, and 6.3 for the strongest single-surrogate baselines. On parking, \sys significantly outperforms every compatible single-surrogate baseline under each search algorithm ($p \le 0.009$). On the humanoid and DonkeyCar, its best means are higher, although several same-search comparisons are not significant.

\textbf{Non-surrogate approaches.} The non-surrogate approaches are substantially less effective. Their best means are 5.7 on parking, 4.0 on the humanoid, and 0.7 on DonkeyCar, versus 14.5, 21.5, and 12.2 for the best ensembled \sys method. \mdpfuzz, \mdpfuzz$^{-}$, and random testing find fewer than one novel failure on parking and DonkeyCar and at most four on the humanoid. Training-failure replay yields zero novel failures by definition because every replay matches a configuration already recorded as failing in the training log. We report failures per simulator validation rather than per unit of time (\Cref{sec:discussion:cost}).

\textbf{Influence of search algorithms.} No single search algorithm dominates across environments. With the ensembled \sys, hill climbing performs best on parking (14.5 novel failures) versus the GA (10.1) and ranking (6.6; $p \le 0.018$). On the humanoid, ranking, the GA, and hill climbing are comparable at 21.5, 21.4, and 19.9 ($p \ge 0.70$). On DonkeyCar, ranking is best at 12.2 versus the GA (3.8) and hill climbing (4.7; $p = 0.001$). The single \sys follows the same pattern: local search is strongest on parking, whereas ranking is strongest on DonkeyCar. On DonkeyCar, the seeded searches spend much of their validation budget replaying known failures, which limits the number of novel failures they find.

\rqconclusion{RQ2}{The ensembled \sys combined with search is the most effective test-generation method in every environment, finding 29\%, 22\%, and 79\% more novel failures than the strongest competing method on parking, Humanoid, and DonkeyCar. This advantage is significant on parking and DonkeyCar and statistically comparable on Humanoid. A single \sys also attains the highest mean among single-surrogate methods, whereas non-surrogate approaches find substantially fewer novel failures. The best search algorithm depends on the environment.}

\subsection{RQ3: Failure Diversity}
\label{sec:results:rq3}

\begin{figure}[tb]
  \centering
  \includegraphics[width=0.8\textwidth]{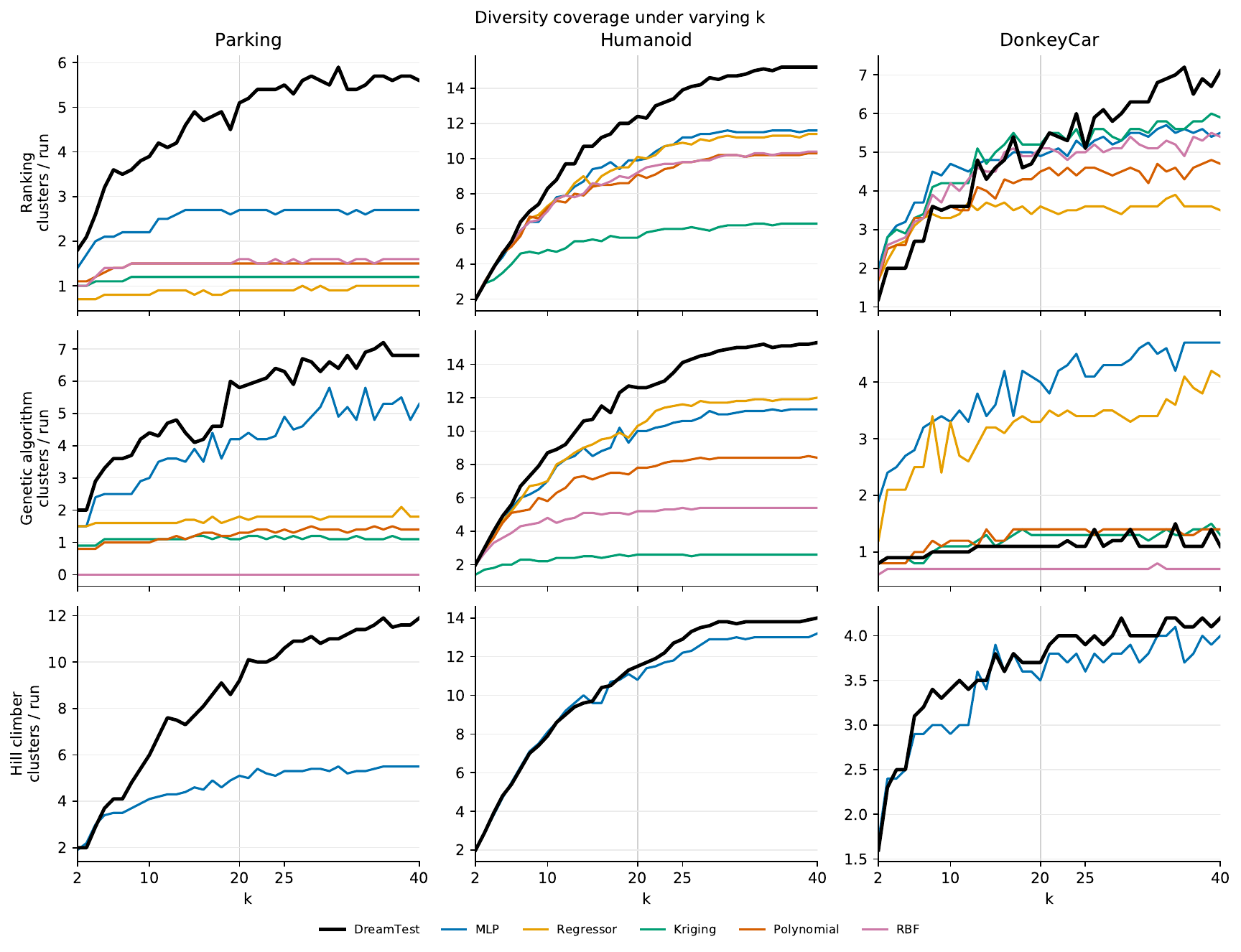}
  \caption{RQ3, failure diversity: mean number of behavioural clusters covered per run over ten runs under the fixed budget of 50 simulator validations, as the number of clusters $k$ varies from 2 to 40. Higher values indicate broader behavioural coverage under the same testing budget.}
  \label{fig:coverage-vs-k}
\end{figure}

\textbf{Comparison across $k$.} \Cref{fig:coverage-vs-k} reports the number of behavioural clusters covered for each $k\in\{2,\ldots,40\}$, rather than for a single value chosen using the silhouette coefficient. We therefore compare methods separately at each $k$. \sys covers the most clusters for almost every $k$ under ranking, and for most $k$ under hill climbing on Parking and the genetic algorithm on Parking and Humanoid. This advantage is thus not an artefact of selecting a favourable $k$.

\noindent\textbf{Search algorithms and environments.} Under ranking, \sys covers the most clusters across almost the entire range on Parking and Humanoid. On DonkeyCar, several baselines cover more clusters at small $k$, whereas \sys leads at larger $k$, where the clustering represents behavioural differences more finely.  Under the genetic algorithm, \sys is again clearly strongest on Parking and Humanoid, with DonkeyCar as the main exception. Under hill climbing, where only \sys and the MLP apply, \sys has a large and persistent advantage on Parking. Results are closer on Humanoid and DonkeyCar, with occasional crossings, but \sys typically covers more clusters as the behavioural representation becomes more fine-grained at larger $k$.  Overall, the largest and most consistent diversity gains occur on Parking; Humanoid also shows clear gains except under hill climbing. DonkeyCar is more sensitive to the search algorithm, with a clear advantage under ranking, comparable performance under hill climbing, and weaker results under the genetic algorithm.

\noindent\textbf{DonkeyCar GA exception.} Under the GA, \sys finds 3.8 novel failures and 1.1 clusters per run on DonkeyCar, compared with 12.2 failures and 4.8 clusters for the same ensemble under ranking. Of its 36.8 raw failures, 33.0 are exact replays. One plausible explanation is that \sys precisely identifies the dense failure region around a seeded road. While this precision is usually beneficial, here it encourages the genetic search to remain in that region, where small mutations repeatedly reproduce the same crash. In contrast, less accurate surrogates yield noisier scores, which may help the search drift away from this basin and uncover more distinct failures. Ranking similarly avoids this local trap and produces substantially more diverse failures. In practice, highly accurate surrogates may need to be paired with alternative search strategies (e.g., ranking or hill climbing) when genetic search becomes trapped in dense local neighbourhoods.

\rqconclusion{RQ3}{Across $k=2,\ldots,40$, the three search algorithms, and all three environments, \sys generally covers more failure-behaviour clusters under the same 50-validation budget than the baselines. The exception is DonkeyCar under the genetic algorithm, where less accurate surrogates cover more behavioural clusters because their less consistent predictions help the local search escape dense regions of known failures.}
\subsection{Ablation Study}
\label{sec:results:ablation}

We ablate \sys’s main design choices to assess its contribution from the world-model rollout, the failure-scoring rule, and ensembling. \Cref{tab:ablation} first removes imagination while retaining the same trajectory-trained RSSM, then varies the imagination horizon and scoring rule.

\begin{table}[H]
  \centering
  \caption{Ablation study of \sys, reported as mean in-distribution historical \auprc\ $\pm$ standard deviation over the ten splits of \Cref{tab:rq1:id}. $H$ denotes the full imagination horizon. Bold marks the highest mean in each column. $^{*}$: significantly different from the full \sys under the two-sided Wilcoxon signed-rank test ($p<0.05$).}
  \label{tab:ablation}
  \scriptsize
  \renewcommand{\arraystretch}{0.95}
  \setlength{\tabcolsep}{4pt}
  \begin{tabular}{@{}llccc@{}}
    \toprule
    Component & Variant & Parking & Humanoid & DonkeyCar \\
    \midrule
    Full model & \sys   & \textbf{0.156$\pm$0.035} & \textbf{0.210$\pm$0.030} & \textbf{0.393$\pm$0.037} \\
    \midrule
    Without imagination & direct classifier on RSSM configuration embedding & 0.121$\pm$0.037$^{*}$ & 0.205$\pm$0.025 & 0.177$\pm$0.019$^{*}$ \\
    \midrule
    \multirow{4}{*}{Imagination horizon}
      & $0$ (no imagined step) & 0.032$\pm$0.009$^{*}$ & 0.073$\pm$0.022$^{*}$ & 0.131$\pm$0.021$^{*}$ \\
      & $1$ step & 0.041$\pm$0.012$^{*}$ & 0.080$\pm$0.021$^{*}$ & 0.111$\pm$0.010$^{*}$ \\
      & $H/4$ & 0.051$\pm$0.010$^{*}$ & 0.107$\pm$0.036$^{*}$ & 0.154$\pm$0.032$^{*}$ \\
      & $H/2$ & 0.126$\pm$0.039$^{*}$ & 0.115$\pm$0.033$^{*}$ & 0.293$\pm$0.046$^{*}$ \\
    \midrule
    \multirow{4}{*}{Scoring rule}
      & average prediction across imagined steps & 0.127$\pm$0.052$^{*}$ & 0.164$\pm$0.027$^{*}$ & 0.155$\pm$0.026$^{*}$ \\
      & highest prediction across imagined steps & 0.039$\pm$0.008$^{*}$ & 0.191$\pm$0.040$^{*}$ & 0.217$\pm$0.067$^{*}$ \\
      & prediction at the final imagined step & 0.062$\pm$0.031$^{*}$ & 0.178$\pm$0.039$^{*}$ & 0.150$\pm$0.023$^{*}$ \\
      & termination-weighted failure score, one rollout & 0.121$\pm$0.034$^{*}$ & 0.187$\pm$0.028$^{*}$ & 0.283$\pm$0.039$^{*}$ \\
    \bottomrule
  \end{tabular}
\end{table}

\textbf{Without imagination.} A key contribution of \sys is its ability to exploit the per-step interaction records already present in the agent's training log. Configuration-only baselines cannot use these records without fundamentally changing their approach. \sys uses the interaction data to learn agent--environment dynamics and then imagines how a test will unfold. This ablation isolates the benefit of imagining the episode. We retain the same RSSM, trained on the same complete trajectories, and freeze its learned parameters. Rather than rolling the learned dynamics forward, we train a linear failure classifier on the RSSM representation of the configuration. Both variants therefore learn from the same per-step records and use the same representation; only the full \sys imagines an episode at inference time.

Without imagination, \auprc\ drops from $0.156\pm0.035$ to $0.121\pm0.037$ on Parking and from $0.393\pm0.037$ to $0.177\pm0.019$ on DonkeyCar; both decreases are significant ($p=0.049$ and $p\le0.002$). The large DonkeyCar drop indicates that a trajectory-trained representation alone is insufficient when failure depends on the agent's interaction with the road. On Humanoid, the direct classifier achieves $0.205\pm0.025$ versus $0.210\pm0.030$ with imagination, with no significant difference. A likely reason is that the Humanoid configuration specifies the agent's initial physical state (e.g., torso pose, joint angles, velocities), which largely determines whether it remains balanced. In contrast, Parking and DonkeyCar outcomes depend more on extended agent--environment interaction, so imagining the episode provides additional predictive signal. Overall, this control shows that the gains on Parking and DonkeyCar are not merely due to training on trajectories, but to using the learned dynamics to imagine episodes at inference time.

\textbf{Imagination horizon.} We shorten the imagined episode while keeping the trained model and scoring procedure fixed. At horizon $0$, \sys scores a configuration without imagining any agent--environment interaction; at horizon $1$, it imagines only the first step. Both variants perform poorly across all three environments. Increasing the horizon to $H/4$ and $H/2$ improves ranking, but every shortened horizon remains significantly worse than the full horizon $H$. At $H/2$, \auprc\ is 0.126, 0.115, and 0.293, compared with 0.156, 0.210, and 0.393 at $H$. Thus, predictions become more informative as more interaction is imagined, with the full episode providing the strongest ranking.

\textbf{Failure-scoring rule.} \sys computes a configuration's failure score by weighting the failure prediction at each imagined step by the probability that the episode terminates at that step, then averaging across $K$ imagined rollouts (\Cref{eq:pfail}). This rule matters in all three environments: replacing it with the temporal average, maximum, or final-step prediction significantly reduces \auprc, with the largest drops on Parking and DonkeyCar. On Humanoid, where 93\% of episodes reach the time limit, the alternatives are closer to the full score but remain significantly worse; the best alternative is the maximum prediction ($0.191\pm0.040$ vs. $0.210\pm0.030$). Keeping the same score but reducing $K$ to one rollout is also significantly worse in every environment, lowering \auprc\ by 0.035, 0.023, and 0.110, indicating that averaging over imagined outcomes is most important on DonkeyCar.

Overall, the ablations identify imagination and the termination-weighted, multi-rollout failure score as the main contributors on Parking and DonkeyCar. The no-imagination RSSM control shows that these gains cannot be explained solely by access to trajectory data. Humanoid depends more on the initial representation, while ensembling improves search stability.

\section{Discussion}
\label{sec:discussion}

\subsection{How Faithful Must the Imagined Episodes Be?}
\label{sec:discussion:fidelity}

RQ1 and the ablation study show that imagined episodes are informative for ranking failures, particularly on parking and DonkeyCar. We therefore assess their fidelity along three dimensions. \textit{Action fidelity} measures whether the model reproduces the actions of the agent under test. \textit{Trajectory fidelity} measures whether the imagined state evolution tracks the real execution. \textit{Termination fidelity} measures whether the imagined episode ends at approximately the correct time. Because \sys uses these episodes to prioritize configurations for simulator validation, the key question is whether they preserve outcome-relevant behavior, not whether they match every simulator state exactly (\Cref{sec:background:wm}).

\textit{Action fidelity.} On in-distribution test episodes (averaged over ten models), the action head reproduces the agent's actions with $R^2 = 0.82$, $0.79$, and $0.64$ on Parking, Humanoid, and DonkeyCar. Thus, the imagined episodes largely match the agent's decisions, although prediction is less accurate on DonkeyCar, where the policy acts from camera images.

\textit{Trajectory fidelity.} After 30 steps, the root mean square error (RMSE) of the mean imagined trajectory is 0.62, 0.48, and 0.55 in standardised units. These errors are lower than those of a persistence predictor (1.29, 1.96, and 1.03) and the mean training trajectory (0.91, 0.53, and 0.60). The model therefore predicts the direction of execution better than these two simple references, although state error still accumulates with the imagination horizon~\cite{janner2019mbpo}.

\textit{Termination fidelity.} Imagined and real episode lengths have Spearman correlations of $\rho = 0.74$ on parking and $0.82$ on DonkeyCar. The correlation is lower on Humanoid ($0.23$), but 93\% of episodes hit the time limit, leaving little duration variation to capture. On DonkeyCar, the termination head can allow an imagined car to continue beyond the real crash; this error compresses failure scores toward zero but can still preserve their ordering.

Overall, this level of fidelity is sufficient for \sys. The surrogate need not reproduce every state or predict the exact failure time; it only needs to rank failure-prone configurations ahead of less informative ones. Useful rankings can be obtained even when trajectories drift and termination is imperfect, making the approach practical. The approximate world model performs inexpensive prioritisation, while the simulator remains the oracle that confirms each outcome. Because failure episodes are oversampled during training, the numerical value of $\pfail$ need not be a calibrated probability~\cite{guo2017calibration}; we use only its ordering.

\subsection{Computational Cost and Practical Use}
\label{sec:discussion:cost}

\sys generates potential failures without executing every candidate in the simulator or real system. The search evaluates candidates using the surrogate and sends only the configurations predicted to fail for validation. This avoids spending costly executions on configurations that are unlikely to reveal failures. The saving is especially valuable for high-fidelity simulation, physical robot testing, and autonomous-vehicle testing, where each run may require substantial computation, equipment, time, or human supervision. Test generation runs offline after agent training, so \sys adds no latency to the deployed agent. We measure training, inference, and end-to-end search time to assess its practicality for offline testing.

Training a single world model takes about 20 minutes on Parking and Humanoid and about one hour on DonkeyCar, compared with less than 10 minutes for the tuned MLP. The ten-model ensemble multiplies this one-time cost per agent, but can be reused across testing runs. Inference is comparatively cheap: batched GPU scoring takes about 1\,ms per configuration and model (about 10\,ms for the ensemble), and ranking a pool of 3\,000 configurations takes about 30\,s.

We also measure end-to-end search efficiency over 50 proposed configurations. Across all environments and search algorithms, \sys confirms the first failure after a median of 1.5--12.5 simulator executions. Competing surrogates require between 1 and 48 executions, and several combinations find no failure within the 50-execution budget. \sys reaches the first failure earlier than other surrogate-guided methods in every parking search. On DonkeyCar, ranking and hill climbing require 2 and 7 executions, compared with 6.5 and 10 for the MLP; results are closer on Humanoid, where most methods find a failure within the first few executions.

A complete 50-execution run with \sys takes 2--30 minutes under ranking across the three environments. The GA takes 25 minutes to two hours, and hill climbing takes 30--60 minutes on Parking and Humanoid. DonkeyCar hill climbing is the exception, taking 30--40 hours for both \sys and the MLP. The bottleneck is not world-model inference: the saliency-guided search repeatedly fails to construct a valid neighbouring road from failure seeds near the generator's validity boundary. Ranking avoids this issue and completes the DonkeyCar run in about 25--30 minutes, making it the practical choice for that environment.

Overall, the added offline cost is acceptable given the benefits: \sys predicts failures more accurately, finds more failures under the same 50-execution budget, and covers more behavioural clusters in most search settings. The resulting larger and more diverse failure set can support more effective agent retraining, although we do not evaluate retraining in this study.

\subsection{Threats to Validity}
\label{sec:discussion:threats}

\textit{Construct validity.} In RQ1, converting each surrogate's score into a pass/fail prediction would require choosing a classification threshold, which could change the comparison. We therefore report \auprc\ and \auroc\ across all thresholds and interpret $\pfail$ only as a ranking score. Diversity similarly depends on the number of clusters, so we report the coverage curve for every $k$ from 2 to 40 using the same clustering procedure for all techniques. These choices reduce metric-selection bias, but novelty and behavioural clusters remain proxies for distinct faults and root causes.

\textit{Internal validity.} \sys uses per-step records, while the baseline surrogates use only configurations and outcomes. This difference is the intended contribution. Existing configuration-only surrogates discard the interaction sequence and cannot exploit it without changing their approaches. The no-imagination and imagination-horizon ablations retain the trajectory-trained RSSM while removing or shortening its rollout. They show that the rollout itself contributes substantially to Parking and DonkeyCar.

Seeded searches can replay logged configurations, so we exclude exact replays from the novel-failure count, report their frequency, and compare surrogates within each fixed search algorithm. Historical labels can also be stale, and stochastic outcomes can change. We therefore report historical and fresh RQ1 labels, apply the same validation protocol to every method, and use ten paired repetitions. Single-execution labels and failure counts nevertheless remain noisy.

\textit{External validity.} Our results may not generalise beyond the evaluated systems. We mitigate this threat by studying three agents from distinct domains and tasks: parking with discrete scene layouts, humanoid locomotion with continuous initial states, and vision-based driving on procedurally generated roads. \sys requires only per-step records already produced during agent training and makes no assumptions specific to these agents. It can therefore be applied to other agents with such logs without additional simulator executions, although broader evaluation is needed to establish generality.

\textit{Conclusion validity.} Randomness in training, search, and simulator execution can affect the results. We mitigate this threat by repeating each experiment ten times and applying the two-sided Wilcoxon signed-rank test. 

\section{Related Work}
\label{sec:related}

\noindent\textit{Testing CPS and DRL agents.} Search-based testing of learning-enabled CPS predates DRL~\cite{abdessalem2016ase,abdessalem2018icse,gambi2019asfault,humeniuk2023ambiegen}, and surveys map the field~\cite{riccio2020mapping,zhang2022mltesting}. For DRL agents, Uesato et al.\ proposed learning a failure predictor from the agent's training history to prioritise which configurations to run~\cite{uesato2019rigorous}. \indago developed this idea into a surrogate-assisted search pipeline~\cite{biagiola2024indago}, and \nexus~\cite{bartlett2025nexus} extended it with a many-objective search algorithm. A later version of \nexus\ evaluates five configuration-based surrogate architectures on the same three subjects and finds the best one to be environment-specific~\cite{bartlett2026reloaded}. Other work searches the agent's state or action space~\cite{zolfagharian2023starla,tappler2022sbtrl}, fuzzes initial states without a model~\cite{pang2022mdpfuzz,mazouni2024mdpfuzzrep,mazouni2024faultdiversity}, or learns a configuration generator online~\cite{lu2022deepcollision}. Most surrogates in this line predict outcomes directly from the static configuration; \sys instead rolls out an imagined episode at query time while leaving the search interface unchanged.

\noindent\textit{Surrogate models in CPS testing.} Surrogate-assisted optimisation~\cite{jin2011surrogate} and CPS falsification have long used Gaussian-process surrogates to reduce expensive simulations~\cite{tuncali2018sim}. \samota brought classical surrogate families to the online testing of driving systems~\cite{haq2022samota,haq2023morlot}. Approximation-based test generation (ARIsTEO) is the closest software-engineering ancestor of learning dynamics before searching: it approximates a compute-intensive CPS model via system identification and searches the approximation~\cite{menghi2020arist}. \sys instead learns the rollout of a learned controller together with its environment, fitted offline to the DRL agent's training log.

\noindent\textit{World models and learned simulators.} Recurrent world models were introduced for policy evolution~\cite{ha2018worldmodels}, PlaNet made the RSSM a standard latent dynamics model~\cite{hafner2019planet}, and Dreamer trains agents inside imagination~\cite{hafner2020dreamer,hafner2025dreamerv3}. Ensembles underpin PETS~\cite{chua2018pets} and model-based policy optimization (MBPO)~\cite{janner2019mbpo}, while value equivalence formalises what a model must get right~\cite{grimm2020valueequiv,lambert2020mismatch}. GAIA-1) synthesises driving video~\cite{hu2023gaia1}, and software engineering has studied learned or lower-fidelity simulators as test oracles~\cite{stocco2023mindgap,stocco2024siblings}.

To our knowledge, \sys is the first world-model surrogate for testing DRL agents. It retains the configuration-scoring interface of prior surrogates, enabling use with existing search algorithms. Unlike configuration-only surrogates, it learns from per-step training records and imagines the agent's interaction with the environment at query time. It also differs from online CPS surrogates and learned simulator oracles: it trains offline on existing logs and validates every selected configuration in the original simulator.

\section{Conclusion and Future Work}
\label{sec:conclusion}

\sys is a surrogate for search-based testing of DRL agents: it learns a world model from the agent's training log and scores a configuration by imagining the episode it would induce. Across the three benchmarks, it ranks failing configurations more accurately than five baseline surrogates on Parking and DonkeyCar, achieves the highest accuracy Humanoid (with the polynomial close behind), and leads on five out-of-distribution configuration sets. Combined with three search algorithms under a 50-execution budget, it finds more novel failures under ranking in every environment, under hill climbing on Parking, and under the genetic algorithm on Humanoid; its best-performing setting also surpasses the strongest baseline on Parking and DonkeyCar.  In the failure-diversity analysis, it covers the most behavioural clusters for almost every $k \in [2,40]$, so the result does not hinge on a single value selected by the silhouette coefficient. When the seeds dominate a seeded search (as on DonkeyCar), we detect no differences between surrogates; when the configuration is the agent's initial state (as on humanoid), we detect no difference relative to the MLP within the gradient-based search. Future work will explore novelty-aware search, querying the real policy instead of a behaviour-cloned copy, and turning imagined episodes into readable failure reports.

\bibliographystyle{unsrtnat}
\bibliography{references}

@inproceedings{abdessalem2016ase, series={ASE'16}, title={Testing advanced driver assistance systems using multi-objective search and neural networks}, url={http://dx.doi.org/10.1145/2970276.2970311}, DOI={10.1145/2970276.2970311}, booktitle={Proceedings of the 31st IEEE/ACM International Conference on Automated Software Engineering}, publisher={ACM}, author={Ben Abdessalem, Raja and Nejati, Shiva and Briand, Lionel C. and Stifter, Thomas}, year={2016}, month={Aug}, pages={63–74}, collection={ASE'16} }

@inproceedings{abdessalem2018icse, series={ICSE ’18}, title={Testing vision-based control systems using learnable evolutionary algorithms}, url={http://dx.doi.org/10.1145/3180155.3180160}, DOI={10.1145/3180155.3180160}, booktitle={Proceedings of the 40th International Conference on Software Engineering}, publisher={ACM}, author={Abdessalem, Raja Ben and Nejati, Shiva and Briand, Lionel C. and Stifter, Thomas}, year={2018}, month={May}, pages={1016–1026}, collection={ICSE ’18} }

@misc{andrychowicz2017her,
  title = {Hindsight Experience Replay},
  author = {Marcin Andrychowicz and Filip Wolski and Alex Ray and Jonas Schneider and Rachel Fong and Peter Welinder and Bob McGrew and Josh Tobin and Pieter Abbeel and Wojciech Zaremba},
  year = {2017},
  eprint = {1707.01495},
  archivePrefix = {arXiv},
  howpublished = {arXiv:1707.01495}
}

@article{arcuri2014hitchhiker, title={A Hitchhiker’s guide to statistical tests for assessing randomized algorithms in software engineering}, volume={24}, ISSN={0960-0833}, url={http://dx.doi.org/10.1002/stvr.1486}, DOI={10.1002/stvr.1486}, number={3}, journal={Software Testing, Verification and Reliability}, publisher={Wiley}, author={Arcuri, Andrea and Briand, Lionel}, year={2012}, month={Nov}, pages={219–250} }

@misc{bartlett2025nexus,
  title = {The Pursuit of Diversity: Multi-Objective Testing of Deep Reinforcement Learning Agents},
  author = {Antony Bartlett and Cynthia Liem and Annibale Panichella},
  year = {2025},
  eprint = {2510.14727},
  archivePrefix = {arXiv},
  howpublished = {arXiv:2510.14727}
}

@article{biagiola2024indago, title={Testing of Deep Reinforcement Learning Agents with Surrogate Models}, volume={33}, ISSN={1557-7392}, url={http://dx.doi.org/10.1145/3631970}, DOI={10.1145/3631970}, number={3}, journal={ACM Transactions on Software Engineering and Methodology}, publisher={Association for Computing Machinery (ACM)}, author={Biagiola, Matteo and Tonella, Paolo}, year={2024}, month={Mar}, pages={1–33} }

@misc{chua2018pets,
  title = {Deep Reinforcement Learning in a Handful of Trials using Probabilistic Dynamics Models},
  author = {Kurtland Chua and Roberto Calandra and Rowan McAllister and Sergey Levine},
  year = {2018},
  eprint = {1805.12114},
  archivePrefix = {arXiv},
  howpublished = {arXiv:1805.12114}
}

@inproceedings{davis2006prroc, series={ICML ’06}, title={The relationship between Precision-Recall and ROC curves}, url={http://dx.doi.org/10.1145/1143844.1143874}, DOI={10.1145/1143844.1143874}, booktitle={Proceedings of the 23rd international conference on Machine learning  - ICML ’06}, publisher={ACM Press}, author={Davis, Jesse and Goadrich, Mark}, year={2006}, pages={233–240}, collection={ICML ’06} }

@misc{gal2016dropout,
  title = {Dropout as a Bayesian Approximation: Representing Model Uncertainty in Deep Learning},
  author = {Yarin Gal and Zoubin Ghahramani},
  year = {2015},
  eprint = {1506.02142},
  archivePrefix = {arXiv},
  howpublished = {arXiv:1506.02142}
}

@article{goel2007ensemble,
  title = {Ensemble of Surrogates},
  author = {Goel, Tushar and Haftka, Raphael T. and Shyy, Wei and Queipo, Nestor V.},
  journal = {Structural and Multidisciplinary Optimization},
  volume = {33},
  number = {3},
  pages = {199--216},
  year = {2007},
  doi = {10.1007/s00158-006-0051-9}
}

@inproceedings{gambi2019asfault, series={ISSTA ’19}, title={Automatically testing self-driving cars with search-based procedural content generation}, url={http://dx.doi.org/10.1145/3293882.3330566}, DOI={10.1145/3293882.3330566}, booktitle={Proceedings of the 28th ACM SIGSOFT International Symposium on Software Testing and Analysis}, publisher={ACM}, author={Gambi, Alessio and Mueller, Marc and Fraser, Gordon}, year={2019}, month={July}, pages={318–328}, collection={ISSTA ’19} }

@misc{grimm2020valueequiv,
  title = {The Value Equivalence Principle for Model-Based Reinforcement Learning},
  author = {Christopher Grimm and André Barreto and Satinder Singh and David Silver},
  year = {2020},
  eprint = {2011.03506},
  archivePrefix = {arXiv},
  howpublished = {arXiv:2011.03506}
}

@misc{guo2017calibration,
  title = {On Calibration of Modern Neural Networks},
  author = {Chuan Guo and Geoff Pleiss and Yu Sun and Kilian Q. Weinberger},
  year = {2017},
  eprint = {1706.04599},
  archivePrefix = {arXiv},
  howpublished = {arXiv:1706.04599}
}

@misc{ha2018worldmodels,
  title = {World Models},
  author = {David Ha and Jürgen Schmidhuber},
  year = {2018},
  eprint = {1803.10122},
  archivePrefix = {arXiv},
  howpublished = {arXiv:1803.10122}
}

@misc{haarnoja2018sac,
  title = {Soft Actor-Critic: Off-Policy Maximum Entropy Deep Reinforcement Learning with a Stochastic Actor},
  author = {Tuomas Haarnoja and Aurick Zhou and Pieter Abbeel and Sergey Levine},
  year = {2018},
  eprint = {1801.01290},
  archivePrefix = {arXiv},
  howpublished = {arXiv:1801.01290}
}

@misc{hafner2019planet,
  title = {Learning Latent Dynamics for Planning from Pixels},
  author = {Danijar Hafner and Timothy Lillicrap and Ian Fischer and Ruben Villegas and David Ha and Honglak Lee and James Davidson},
  year = {2018},
  eprint = {1811.04551},
  archivePrefix = {arXiv},
  howpublished = {arXiv:1811.04551}
}

@misc{hafner2020dreamer,
  title = {Dream to Control: Learning Behaviors by Latent Imagination},
  author = {Danijar Hafner and Timothy Lillicrap and Jimmy Ba and Mohammad Norouzi},
  year = {2019},
  eprint = {1912.01603},
  archivePrefix = {arXiv},
  howpublished = {arXiv:1912.01603}
}

@article{hafner2025dreamerv3, title={Mastering diverse control tasks through world models}, volume={640}, ISSN={1476-4687}, url={http://dx.doi.org/10.1038/s41586-025-08744-2}, DOI={10.1038/s41586-025-08744-2}, number={8059}, journal={Nature}, publisher={Springer Science and Business Media LLC}, author={Hafner, Danijar and Pasukonis, Jurgis and Ba, Jimmy and Lillicrap, Timothy}, year={2025}, month={Apr}, pages={647–653} }

@inproceedings{haq2022samota, series={ICSE ’22}, title={Efficient online testing for DNN-enabled systems using surrogate-assisted and many-objective optimization}, url={http://dx.doi.org/10.1145/3510003.3510188}, DOI={10.1145/3510003.3510188}, booktitle={Proceedings of the 44th International Conference on Software Engineering}, publisher={ACM}, author={Haq, Fitash Ul and Shin, Donghwan and Briand, Lionel}, year={2022}, month={May}, pages={811–822}, collection={ICSE ’22} }

@inproceedings{haq2023morlot, title={Many-Objective Reinforcement Learning for Online Testing of DNN-Enabled Systems}, url={http://dx.doi.org/10.1109/icse48619.2023.00155}, DOI={10.1109/icse48619.2023.00155}, booktitle={2023 IEEE/ACM 45th International Conference on Software Engineering (ICSE)}, publisher={IEEE}, author={Ul Haq, Fitash and Shin, Donghwan and Briand, Lionel C.}, year={2023}, month={May}, pages={1814–1826} }

@misc{hu2023gaia1,
  title = {GAIA-1: A Generative World Model for Autonomous Driving},
  author = {Anthony Hu and Lloyd Russell and Hudson Yeo and Zak Murez and George Fedoseev and Alex Kendall and Jamie Shotton and Gianluca Corrado},
  year = {2023},
  eprint = {2309.17080},
  archivePrefix = {arXiv},
  howpublished = {arXiv:2309.17080}
}

@article{humeniuk2023ambiegen, title={AmbieGen: A search-based framework for autonomous systems testing}, volume={230}, ISSN={0167-6423}, url={http://dx.doi.org/10.1016/j.scico.2023.102990}, DOI={10.1016/j.scico.2023.102990}, journal={Science of Computer Programming}, publisher={Elsevier BV}, author={Humeniuk, Dmytro and Khomh, Foutse and Antoniol, Giuliano}, year={2023}, month={Aug}, pages={102990} }

@misc{janner2019mbpo,
  title = {When to Trust Your Model: Model-Based Policy Optimization},
  author = {Michael Janner and Justin Fu and Marvin Zhang and Sergey Levine},
  year = {2019},
  eprint = {1906.08253},
  archivePrefix = {arXiv},
  howpublished = {arXiv:1906.08253}
}

@article{jin2011surrogate, title={Surrogate-assisted evolutionary computation: Recent advances and future challenges}, volume={1}, ISSN={2210-6502}, url={http://dx.doi.org/10.1016/j.swevo.2011.05.001}, DOI={10.1016/j.swevo.2011.05.001}, number={2}, journal={Swarm and Evolutionary Computation}, publisher={Elsevier BV}, author={Jin, Yaochu}, year={2011}, month={June}, pages={61–70} }

@misc{kuznetsov2020tqc,
  title = {Controlling Overestimation Bias with Truncated Mixture of Continuous Distributional Quantile Critics},
  author = {Arsenii Kuznetsov and Pavel Shvechikov and Alexander Grishin and Dmitry Vetrov},
  year = {2020},
  eprint = {2005.04269},
  archivePrefix = {arXiv},
  howpublished = {arXiv:2005.04269}
}

@misc{lakshminarayanan2017ensembles,
  title = {Simple and Scalable Predictive Uncertainty Estimation using Deep Ensembles},
  author = {Balaji Lakshminarayanan and Alexander Pritzel and Charles Blundell},
  year = {2016},
  eprint = {1612.01474},
  archivePrefix = {arXiv},
  howpublished = {arXiv:1612.01474}
}

@misc{lambert2020mismatch,
  title = {Objective Mismatch in Model-based Reinforcement Learning},
  author = {Nathan Lambert and Brandon Amos and Omry Yadan and Roberto Calandra},
  year = {2020},
  eprint = {2002.04523},
  archivePrefix = {arXiv},
  howpublished = {arXiv:2002.04523}
}

@article{lu2022deepcollision, title={Learning Configurations of Operating Environment of Autonomous Vehicles to Maximize their Collisions}, volume={49}, ISSN={2326-3881}, url={http://dx.doi.org/10.1109/TSE.2022.3150788}, DOI={10.1109/tse.2022.3150788}, number={1}, journal={IEEE Transactions on Software Engineering}, publisher={Institute of Electrical and Electronics Engineers (IEEE)}, author={Lu, Chengjie and Shi, Yize and Zhang, Huihui and Zhang, Man and Wang, Tiexin and Yue, Tao and Ali, Shaukat}, year={2023}, month={Jan}, pages={384–402} }

@misc{mazouni2024faultdiversity,
  title = {Testing for Fault Diversity in Reinforcement Learning},
  author = {Quentin Mazouni and Helge Spieker and Arnaud Gotlieb and Mathieu Acher},
  year = {2024},
  eprint = {2403.15065},
  archivePrefix = {arXiv},
  howpublished = {arXiv:2403.15065}
}

@inproceedings{menghi2020arist, series={ICSE ’20}, title={Approximation-refinement testing of compute-intensive cyber-physical models: an approach based on system identification}, url={http://dx.doi.org/10.1145/3377811.3380370}, DOI={10.1145/3377811.3380370}, booktitle={Proceedings of the ACM/IEEE 42nd International Conference on Software Engineering}, publisher={ACM}, author={Menghi, Claudio and Nejati, Shiva and Briand, Lionel and Parache, Yago Isasi}, year={2020}, month={June}, pages={372–384}, collection={ICSE ’20} }

@inproceedings{pang2022mdpfuzz, series={ISSTA ’22}, title={MDPFuzz: testing models solving Markov decision processes}, url={http://dx.doi.org/10.1145/3533767.3534388}, DOI={10.1145/3533767.3534388}, booktitle={Proceedings of the 31st ACM SIGSOFT International Symposium on Software Testing and Analysis}, publisher={ACM}, author={Pang, Qi and Yuan, Yuanyuan and Wang, Shuai}, year={2022}, month={July}, pages={378–390}, collection={ISSTA ’22} }

@article{riccio2020mapping, title={Testing machine learning based systems: a systematic mapping}, volume={25}, ISSN={1573-7616}, url={http://dx.doi.org/10.1007/s10664-020-09881-0}, DOI={10.1007/s10664-020-09881-0}, number={6}, journal={Empirical Software Engineering}, publisher={Springer Science and Business Media LLC}, author={Riccio, Vincenzo and Jahangirova, Gunel and Stocco, Andrea and Humbatova, Nargiz and Weiss, Michael and Tonella, Paolo}, year={2020}, month={sep}, pages={5193–5254} }

@article{rousseeuw1987silhouette, title={Silhouettes: A graphical aid to the interpretation and validation of cluster analysis}, volume={20}, ISSN={0377-0427}, url={http://dx.doi.org/10.1016/0377-0427(87)90125-7}, DOI={10.1016/0377-0427(87)90125-7}, journal={Journal of Computational and Applied Mathematics}, publisher={Elsevier BV}, author={Rousseeuw, Peter J.}, year={1987}, month={Nov}, pages={53–65} }

@article{saito2015prc, title={The Precision-Recall Plot Is More Informative than the ROC Plot When Evaluating Binary Classifiers on Imbalanced Datasets}, volume={10}, ISSN={1932-6203}, url={http://dx.doi.org/10.1371/journal.pone.0118432}, DOI={10.1371/journal.pone.0118432}, number={3}, journal={PLOS ONE}, publisher={Public Library of Science (PLoS)}, author={Saito, Takaya and Rehmsmeier, Marc}, editor={Brock, Guy}, year={2015}, month={Mar}, pages={e0118432} }

@misc{simonyan2013saliency,
  title = {Deep Inside Convolutional Networks: Visualising Image Classification Models and Saliency Maps},
  author = {Karen Simonyan and Andrea Vedaldi and Andrew Zisserman},
  year = {2013},
  eprint = {1312.6034},
  archivePrefix = {arXiv},
  howpublished = {arXiv:1312.6034}
}

@article{stocco2023mindgap, title={Mind the Gap! A Study on the Transferability of Virtual Versus Physical-World Testing of Autonomous Driving Systems}, volume={49}, ISSN={2326-3881}, url={http://dx.doi.org/10.1109/TSE.2022.3202311}, DOI={10.1109/tse.2022.3202311}, number={4}, journal={IEEE Transactions on Software Engineering}, publisher={Institute of Electrical and Electronics Engineers (IEEE)}, author={Stocco, Andrea and Pulfer, Brian and Tonella, Paolo}, year={2023}, month={Apr}, pages={1928–1940} }

@article{stocco2024siblings, title={Two is better than one: digital siblings to improve autonomous driving testing}, volume={29}, ISSN={1573-7616}, url={http://dx.doi.org/10.1007/s10664-024-10458-4}, DOI={10.1007/s10664-024-10458-4}, number={4}, journal={Empirical Software Engineering}, publisher={Springer Science and Business Media LLC}, author={Biagiola, Matteo and Stocco, Andrea and Riccio, Vincenzo and Tonella, Paolo}, year={2024}, month={May} }

@misc{tappler2022sbtrl,
  title = {Search-Based Testing of Reinforcement Learning},
  author = {Martin Tappler and Filip Cano Córdoba and Bernhard K. Aichernig and Bettina Könighofer},
  year = {2022},
  eprint = {2205.04887},
  archivePrefix = {arXiv},
  howpublished = {arXiv:2205.04887}
}

@inproceedings{todorov2012mujoco, title={MuJoCo: A physics engine for model-based control}, url={http://dx.doi.org/10.1109/IROS.2012.6386109}, DOI={10.1109/iros.2012.6386109}, booktitle={2012 IEEE/RSJ International Conference on Intelligent Robots and Systems}, publisher={IEEE}, author={Todorov, Emanuel and Erez, Tom and Tassa, Yuval}, year={2012}, month={Oct}, pages={5026–5033} }

@inproceedings{tuncali2018sim, title={Simulation-based Adversarial Test Generation for Autonomous Vehicles with Machine Learning Components}, url={http://dx.doi.org/10.1109/IVS.2018.8500421}, DOI={10.1109/ivs.2018.8500421}, booktitle={2018 IEEE Intelligent Vehicles Symposium (IV)}, publisher={IEEE}, author={Tuncali, Cumhur Erkan and Fainekos, Georgios and Ito, Hisahiro and Kapinski, James}, year={2018}, month={June}, pages={1555–1562} }

@misc{uesato2019rigorous,
  title = {Rigorous Agent Evaluation: An Adversarial Approach to Uncover Catastrophic Failures},
  author = {Jonathan Uesato and Ananya Kumar and Csaba Szepesvari and Tom Erez and Avraham Ruderman and Keith Anderson and  Krishmamurthy  Dvijotham and Nicolas Heess and Pushmeet Kohli},
  year = {2018},
  eprint = {1812.01647},
  archivePrefix = {arXiv},
  howpublished = {arXiv:1812.01647}
}

@article{vargha2000a12, title={A Critique and Improvement of the CL Common Language Effect Size Statistics of McGraw and Wong}, volume={25}, ISSN={1935-1054}, url={http://dx.doi.org/10.3102/10769986025002101}, DOI={10.3102/10769986025002101}, number={2}, journal={Journal of Educational and Behavioral Statistics}, publisher={American Educational Research Association (AERA)}, author={Vargha, András and Delaney, Harold D.}, year={2000}, month={June}, pages={101–132} }

@article{zhang2022mltesting, title={Machine Learning Testing: Survey, Landscapes and Horizons}, volume={48}, ISSN={2326-3881}, url={http://dx.doi.org/10.1109/TSE.2019.2962027}, DOI={10.1109/tse.2019.2962027}, number={1}, journal={IEEE Transactions on Software Engineering}, publisher={Institute of Electrical and Electronics Engineers (IEEE)}, author={Zhang, Jie M. and Harman, Mark and Ma, Lei and Liu, Yang}, year={2022}, month={Jan}, pages={1–36} }

@article{zolfagharian2023starla, title={A Search-Based Testing Approach for Deep Reinforcement Learning Agents}, volume={49}, ISSN={2326-3881}, url={http://dx.doi.org/10.1109/TSE.2023.3269804}, DOI={10.1109/tse.2023.3269804}, number={7}, journal={IEEE Transactions on Software Engineering}, publisher={Institute of Electrical and Electronics Engineers (IEEE)}, author={Zolfagharian, Amirhossein and Abdellatif, Manel and Briand, Lionel C. and Bagherzadeh, Mojtaba and S, Ramesh}, year={2023}, month={July}, pages={3715–3735} }

@misc{leurent2018highway,
  title = {An Environment for Autonomous Driving Decision-Making},
  author = {Leurent, Edouard},
  year = {2018},
  howpublished = {\url{https://github.com/eleurent/highway-env}}
}

@misc{donkeysim,
  title = {Donkey Car Simulator},
  author = {{Tawn Kramer and contributors}},
  year = {2019},
  howpublished = {\url{https://github.com/tawnkramer/gym-donkeycar}}
}

@inproceedings{mazouni2024mdpfuzzrep, series={ISSTA ’24}, title={Policy Testing with MDPFuzz (Replicability Study)}, url={http://dx.doi.org/10.1145/3650212.3680382}, DOI={10.1145/3650212.3680382}, booktitle={Proceedings of the 33rd ACM SIGSOFT International Symposium on Software Testing and Analysis}, publisher={ACM}, author={Mazouni, Quentin and Spieker, Helge and Gotlieb, Arnaud and Acher, Mathieu}, year={2024}, month={sep}, pages={1567–1578}, collection={ISSTA ’24} }

@misc{bartlett2026reloaded,
  title = {Surrogate Reloaded: Multi-Objective Search for Diverse Failure Scenarios in Deep Reinforcement Learning Agents},
  author = {Antony Bartlett and Cynthia Liem and Annibale Panichella},
  year = {2026},
  howpublished = {Research Square preprint, rs-9426270/v1, posted 8 May 2026},
  doi = {10.21203/rs.3.rs-9426270/v1}
}

@inproceedings{wang2025qurebot,
  title = {Quantum Machine Learning-based Test Oracle for Autonomous Mobile Robots},
  author = {Wang, Xinyi and Xu, Qinghua and Arcaini, Paolo and Ali, Shaukat and Peyrucain, Thomas},
  booktitle = {2025 40th IEEE/ACM International Conference on Automated Software Engineering (ASE)},
  pages = {3695--3706},
  year = {2025},
  doi = {10.1109/ASE63991.2025.00306}
}

@article{hadadi2026flexlog,
  title = {{LLM} Meets {ML}: Data-efficient Anomaly Detection on Unstable Logs},
  author = {Hadadi, Fatemeh and Xu, Qinghua and Bianculli, Domenico and Briand, Lionel},
  journal = {ACM Transactions on Software Engineering and Methodology},
  volume = {35},
  number = {8},
  pages = {216:1--216:48},
  year = {2026},
  doi = {10.1145/3771283}
}

@article{xu2024ppt,
  title = {Pretrain, Prompt, and Transfer: Evolving Digital Twins for Time-to-Event Analysis in Cyber-Physical Systems},
  author = {Xu, Qinghua and Yue, Tao and Ali, Shaukat and Arratibel, Maite},
  journal = {IEEE Transactions on Software Engineering},
  volume = {50},
  number = {6},
  pages = {1464--1477},
  year = {2024},
  doi = {10.1109/TSE.2024.3388572}
}

@inproceedings{xu2023kddt,
  title = {{KDDT}: Knowledge Distillation-Empowered Digital Twin for Anomaly Detection},
  author = {Xu, Qinghua and Ali, Shaukat and Yue, Tao and {Zaimovic Nedim} and Singh, Inderjeet},
  booktitle = {Proceedings of the 31st ACM Joint European Software Engineering Conference and Symposium on the Foundations of Software Engineering},
  pages = {1867--1878},
  year = {2023},
  doi = {10.1145/3611643.3613879}
}

\end{document}